\documentclass[11pt]{article}
\usepackage{verbatim,color}
\usepackage{amsthm,amsmath,amssymb,amsfonts}
\usepackage{graphicx}
\usepackage{bm}
\usepackage{natbib}
\bibpunct{(}{)}{;}{a}{}{,}
\usepackage[noline, ruled,boxed]{algorithm2e}
\usepackage[tight]{subfigure}
\usepackage{booktabs}
\usepackage{arydshln}  
\usepackage{setspace}
\usepackage{afterpage}

\newcommand{\blind}{0}

\newtheorem{theorem}{Theorem}[section]
\newtheorem{proposition}[theorem]{Proposition}
\newtheorem{lemma}[theorem]{Lemma}
\newtheorem{definition}[theorem]{Definition}

\begin{document}

\def\spacingset#1{\renewcommand{\baselinestretch}%
{#1}\small\normalsize} \spacingset{1}


\if0\blind
{
  \title{\bf Residual Weighted Learning for Estimating Individualized Treatment Rules}
\author{Xin Zhou, Nicole Mayer-Hamblett, Umer Khan and Michael R. Kosorok\thanks{
    Xin Zhou is a PhD student, Department of Biostatistics, University of North Carolina at Chapel Hill, NC 27599 (email: \texttt{xinzhou@live.unc.edu}).
Nicole Mayer-Hamblett is Associate Professor, Department of Pediatrics, and Adjunct Associate Professor, Department of Biostatistics, University of Washington (email:  \texttt{nicole.hamblett@seattlechildrens.org}).
Umer Khan is a statistician, Cystic Fibrosis Foundation Therapeutics Development Network, Seattle Children’s Hospital (email: \texttt{umer.khan@seattlechildrens.org}).
Michael R. Kosorok is W. R. Kenan, Jr. Distiguished Professor and Chair,
Department of Biostatistics, and Professor, Department of Statistics and Operations Research, University of North Carolina at Chapel Hill, Chapel Hill, NC 27599 (email: \texttt{kosorok@unc.edu}).
The EPIC trial was funded by the National Heart Lung and Blood Institute (NHLBI) and National Institute of Diabetes and Digestive and Kidney Diseases (NIDDK) (U01-HL080310) and the Cystic Fibrosis Foundation Therapeutics. The first and fourth authors were funded in part by the CTSA at the University of North Carolina at Chapel Hill (UL1TR001111) as well as by grant P01CA142538 from the National Cancer Institute (NCI). The second and third authors were funded in part by the CTSA at the University of Washington (UL1TR000423).
The authors thank Editor Nicholas Jewell, associate editor and two reviewers for their helpful comments which led to a significantly improved paper.}}
  \maketitle
} \fi

\if1\blind
{
  \bigskip
  \bigskip
  \bigskip
  \begin{center}
    {\LARGE\bf Residual Weighted Learning for Estimating Individualized Treatment Rules}
\end{center}
  \medskip
} \fi

\bigskip
\begin{abstract}
Personalized medicine has received increasing attention among statisticians, computer scientists, and clinical practitioners. A major component of personalized medicine is the estimation of individualized treatment rules (ITRs). Recently, \citet{Zhao:OWL2012} proposed outcome weighted learning (OWL) to construct ITRs that directly optimize the clinical outcome. Although OWL opens the door to introducing machine learning techniques to optimal treatment regimes, it still has some problems in performance. (1) The estimated ITR of OWL is affected by a simple shift of the outcome. (2) The rule from OWL tries to keep treatment assignments that subjects actually received. (3) There is no variable selection mechanism with OWL. All of them weaken the finite sample performance of OWL.
In this article, we propose a general framework, called Residual Weighted Learning (RWL), to alleviate these problems, and hence to improve finite sample performance. Unlike OWL which weights misclassification errors by clinical outcomes, RWL weights these errors by residuals of the outcome from a regression fit on clinical covariates excluding treatment assignment. We utilize the smoothed ramp loss function in RWL, and provide a difference of convex (d.c.) algorithm to solve the corresponding non-convex optimization problem. By estimating residuals with linear models or generalized linear models, RWL can effectively deal with different types of outcomes, such as continuous, binary and count outcomes. We also propose variable selection methods for linear and nonlinear rules, respectively, to further improve the performance.
We show that the resulting estimator of the treatment rule
is consistent. We further obtain a rate of convergence for the difference between the expected
outcome using the estimated ITR and that of the optimal treatment rule.
The performance of the proposed RWL methods is illustrated in simulation studies and in an analysis of cystic fibrosis clinical trial data.
\end{abstract}

\noindent%


\section{Introduction}
Personalized medicine is a medical paradigm that utilizes individual patient information to optimize patients health care. Recently, personalized medicine has received much attention among statisticians, computer scientists, and clinical practitioners. The primary motivation is the well-established fact that patients often show significant heterogeneity in response to treatments. For instance, in a recent study, it was demonstrated that the optimal timing for the initiation of antiretroviral therapy (ART) varies in patients co-infected with human immunodeficiency virus and tuberculosis. Patients with CD4+ T-cell counts of less than 50 per cubic millimeter benefited substantially from earlier ART with a lower rate of new AIDS-defining illnesses and mortality as compared with later ART, while those with larger CD4+ T-cell counts did not have such a benefit \citep{Havlir:ARTTiming2011}. The inherent heterogeneity across patients suggests a transition from the traditional ``one size fits all'' approach to modern personalized medicine.

A major component of personalized medicine is the estimation of individualized treatment rules (ITRs). Formally, the goal is to seek a rule that assigns a treatment, from among a set of possible treatments, to a patient based on his or her clinical, prognostic or genomic characteristics. The individualized treatment rules are also called optimal treatment regimes. There is a significant literature on individualized treatment strategies based on data from clinical trials or observational studies \citep{Murphy:OptimalDTR2003,Robins2004:SNM,Zhang:RobustITR2012, Zhao:RL2009}. Much of the work relies on modeling either the conditional mean outcomes or contrasts between mean outcomes. These methods obtain ITRs indirectly by inverting the regression estimates instead of directly optimizing the decision rule. For instance, \citet{Qian:ITR2011} applied a two-step procedure that first estimates a conditional mean for the outcome and then determines the treatment rule by comparing the conditional means across various treatments. The success of these indirect approaches highly depends on correct specification of posited models and on the precision of model estimates.

In contrast, \citet{Zhao:OWL2012} proposed outcome weighted learning (OWL), using data from a randomized clinical trial, to construct an ITR that directly optimizes the clinical outcome. They cast the treatment selection problem as a weighted classification problem, and apply state-of-the-art support vector machines for implementation. This approach opens the door to introducing machine learning techniques into this area. However, there is still significant room for improved performance. First, the estimated ITR of OWL is affected by a simple shift of the outcome. This behavior makes the estimate of OWL unstable. Second, since OWL requires the outcome to be nonnegative, OWL works similarly to weighted classification, in which misclassification errors, the differences between the estimated and true treatment assignments, are targeted to be reduced. Hence the ITR estimated by OWL tries to keep treatment assignments that subjects actually received. This is not always ideal since treatments are randomly assigned in the trial, and the probability is slim that the majority of subjects are assigned optimal treatments. Third, OWL does not have variable selection features. When there are many clinical covariates which are not related to the heterogeneous treatment effects, variable selection is critical to the performance.

To alleviate these problems, we propose a new method, called Residual Weighted Learning (RWL). Unlike OWL which weights misclassification errors by clinical outcomes, RWL weights these errors by residuals from a regression fit of the outcome. The predictors of the regression model include clinical covariates, but exclude treatment assignment. Thus the residuals better reflect the heterogeneity of treatment effects.
Since some residuals are negative, the hinge loss function used in OWL is not appropriate. We instead utilize the smoothed ramp loss function,
and provide a difference of convex (d.c.) algorithm to solve the corresponding non-convex optimization problem. The smoothed ramp loss resembles the ramp loss \citep{Collobert2006:RampLoss}, but it is smooth everywhere. It is well known that ramp loss related methods are shown to be robust to outliers \citep{Wu2007:RobustSVM}. The robustness to outliers for the smoothed ramp loss is helpful for RWL, especially when residuals are poorly estimated. Moreover, through using residuals, RWL is able to deal relatively easily with almost all types of outcomes. For example, RWL can work with generalized linear models to construct ITRs for count/rate outcomes. We also propose variable selection approaches in RWL for linear and nonlinear rules, respectively, to further improve finite sample performance.

The theoretical analysis of RWL focuses on two aspects, universal consistency and convergence rate. The notion of universal consistency is borrowed from machine learning. It requires for a learning method that when the sample size approaches infinity the method eventually learns the Bayes rule without knowing any specifics of the distribution of the data. We show that RWL with a universal kernel (e.g. Gaussian RBF kernel) is universally consistent.
In machine learning, there is a famous ``no-free-lunch theorem'', which states that the convergence rate of any particular learning rule may be arbitrarily slow \citep{Devroye:PatternRecog1996}. In this article, we prove the ``no-free-lunch theorem'' for finding ITRs. Thus the rate of convergence studies for a particular rule must necessarily be accompanied by conditions on the distribution of the data. For RWL with Gaussian RBF kernel, we show that under the geometric noise condition \citep{Steinwart:FastRateGaussian} the convergence rate is as high as $n^{-1/3}$.

At first glance, one may think that there is not a large difference between OWL and RWL except that RWL uses residuals as alternative outcomes. Actually, RWL enjoys many benefits from this simple modification. First, by using residuals of outcomes, RWL is able to reduce the variability introduced by the original outcomes.
Second, since the numbers of subjects with positive and negative residuals are generally balanced, the ITR determined by RWL favors neither the treatment assignments that subjects actually received nor their opposites. Because of the above reasons, RWL improves finite sample performance. Third, RWL possesses location-scale invariance with respect to the original outcomes. Specifically, the estimated rule of RWL is invariant to a shift of the outcome; it is invariant to a scaling of the outcome with a positive number; the rule from RWL that maximizes the outcome is opposite to the rule that minimizes the outcome. These are intuitively sensible.

The contributions of this article are summarized as follows. (1) We propose the general framework of Residual Weighted Learning to estimate individualized treatment rules. By estimating residuals with linear or generalized linear models, RWL can effectively deal with different types of outcomes, such as continuous, binary and count outcomes. For censored survival outcomes, RWL could potentially utilize martingale residuals, although theoretical justification is still under development. (2) We develop variable selection techniques in RWL to further improve performance. (3) We present a comprehensive theoretical analysis of RWL on universal consistency and convergence rate. (4) As a by-product, we show the ``no-free-lunch theorem'' for ITRs that the convergence rate of any particular rule may be arbitrarily slow. This is a generic result, and it applies to any algorithm for optimal treatment regimes.

The remainder of the article is organized as follows. In Section 2, we discuss Outcome Weighted Learning (OWL) and propose Residual Weighted Learning (RWL) to improve finite sample performance for continuous outcomes. Then we develop a general framework for RWL to handle other types of outcomes, and use binary and count/rate outcomes as examples.
In Section 3, we establish consistency and convergence rate results for the estimated rules. The variable selection techniques for RWL are discussed in Section 4. We present simulation studies to evaluate performance of the proposed methods in Section 5. The method is then illustrated on the EPIC cystic fibrosis randomized clinical trial \citep{Treggiari2009:EPIC, Treggiari2011:EPIC} in Section 6. We conclude the article with a discussion in Section 7. All proofs are given in Appendix.

\section{Methodology}
\subsection{Outcome Weighted Learning}
Consider a two-arm randomized trial. We observe a triplet $(\bm{X}, A, R)$ from each patient, where $\bm{X}=(X_1,\cdots,X_p)^T\in\mathcal{X}$ denotes the patient's clinical covariates, $A\in\mathcal{A}=\{1,-1\}$ denotes the treatment assignment, and $R$ is the observed clinical outcome, also called the ``reward'' in the literature on reinforcement learning. We assume that $R$ is bounded, and larger values of $R$ are more desirable. An individualized treatment rule (ITR) is a function from $\mathcal{X}$ to $\mathcal{A}$. Let $\pi(a,\bm{x}):=P(A=a|\bm{X}=\bm{x})$ be the probability of being assigned treatment $a$ for patients with clinical covariates $\bm{x}$. It is predefined in the trial design. Here we consider a general situation. In most clinical trials, the treatment assignment is independent of $\bm{X}$, but in some designs, such as stratified designs, $A$ may depend on $\bm{X}$. We assume $\pi(a,\bm{x})>0$ for all $a\in\mathcal{A}$ and $\bm{x}\in\mathcal{X}$.

An optimal ITR is a rule that maximizes the expected outcome under this rule. Mathematically, the expected outcome under any ITR $d$ is given as
\begin{equation} \label{eq:ITR}
\mathbb{E}(R|A=d(\bm{X}))=\mathbb{E}\left(\frac{R}{\pi(A,\bm{X})}\mathbb{I}\Big(A=d(\bm{X})\Big)\right),
\end{equation}
where $\mathbb{I}(\cdot)$ is the indicator function. Interested readers may refer to \citet{Qian:ITR2011} and \citet{Zhao:OWL2012} regarding the derivation of (\ref{eq:ITR}). This expectation is called the value function associated with the rule $d$, and is denoted $V(d)$. In other words, the value function $V(d)$ is the expected value of $R$ given that the rule $d(\bm{X})$ is applied to the given population of patients. An optimal ITR $d^*$ is a rule that maximizes $V(d)$. Finding $d^*$ is equivalent to the following minimization problem:
\begin{equation}\label{eq:weightederror}
d^*\in \arg\min_{d}\mathbb{E}\left(\frac{R}{\pi(A,\bm{X})}\mathbb{I}\Big(A\neq d(\bm{X})\Big)\right).
\end{equation}
\citet{Zhao:OWL2012} viewed this as a weighted classification problem, in which one wants to classify $A$ using $\bm{X}$ but also weights each misclassification error by $R/\pi(A,\bm{X})$.

Assume that the observed data $\{(\bm{x}_i, a_i, r_i): i=1,\cdots,n\}$ are collected independently. For any decision function $f(\bm{x})$, let $d_f(\bm{x})=sign\big(f(\bm{x})\big)$ be the associated rule, where $sign(u)=1$ for $u>0$ and $-1$ otherwise. The particular choice of the value of $sign(0)$ is not important. In this article, we fix $sign(0)=-1$. Using the observed data, the weighted classification error in (\ref{eq:weightederror}) can be approximated by the empirical risk
\begin{equation} \label{eq:empiricleweightederror}
\frac{1}{n}\sum_{i=1}^{n}\frac{r_i}{\pi(a_i,\bm{x}_i)}\mathbb{I}\Big(d_f(\bm{x}_i)\neq a_i\Big).
\end{equation}
The optimal decision function $\hat{f}(\bm{x})$ is the one that minimizes the outcome weighted error (\ref{eq:empiricleweightederror}).

It is well known that empirical risk minimization for a classification problem with the 0-1 loss function is an NP-hard problem. To alleviate this difficulty, one often finds a surrogate loss to replace the 0-1 loss. Outcome weighted learning (OWL) proposed by \citet{Zhao:OWL2012} uses the hinge loss function, and also applies the regularization technique used in the support vector machine (SVM) \citep{Vapnik:svm98}. In other words, instead of minimizing (\ref{eq:empiricleweightederror}), OWL aims to minimize
\begin{equation} \label{eq:owloptimization}
\frac{1}{n}\sum_{i=1}^{n}\frac{r_i}{\pi(a_i,\bm{x}_i)}\Big(1-a_if(\bm{x}_i)\Big)_++\lambda||f||^2,
\end{equation}
where $(u)_+=\max(u,0)$ is the positive part of $u$, $||f||$ is some norm for $f$, and $\lambda$ is a tuning parameter controlling the trade-off between empirical risk and complexity of the decision function $f$.

OWL opens the door to the application of statistical learning techniques to personalized medicine. However, this approach is not perfect. First, a simple shift on the outcome $R$ should not affect the optimal decision rule, as seen from (\ref{eq:ITR}) and (\ref{eq:weightederror}). That is, $d^*$ in (\ref{eq:weightederror}) does not change if $R$ is replaced by $R+c$, for any constant $c$. Unfortunately this nice invariance property does not hold for the decision function $\hat{f}(\bm{x})$ of OWL in (\ref{eq:owloptimization}) when $r_i$ is replaced by $r_i+c$. Consider an extreme case where $c$ is very large and $\pi(1,\bm{x})=\pi(-1,\bm{x})=0.5$ for all $\bm{x}\in\mathcal{X}$, the weights $(r_i+c)/\pi(a_i,\bm{x}_i)$ are almost identical, and the weighted problem is approximately transformed to an unweighted one. Hence the performance of OWL can be affected by a simple shift of $R$. Second, OWL further assumes that $R$ is nonnegative to gain computational efficiency from convex programming. A direct consequence of this assumption, as seen from (\ref{eq:empiricleweightederror}), is that the treatment regime $d_f(\bm{x}_i)$ tends to match the treatment $a_i$ that was actually assigned to the patient, especially when the decision function $f$ is chosen from a rich class of functions. This property is not ideal for data from a randomized clinical trial, since treatments are actually randomly assigned to patients.

\subsection{Residual Weighted Learning}\label{sec:rwl}

In this section, we only consider the case that the outcome $R$ is continuous, and extend the framework to other types of outcomes in Section \ref{sec:generalframework}.

As demonstrated previously, the decision rule in (\ref{eq:weightederror}) is invariant to a shift of outcome $R$ by any constant. Moreover, Lemma \ref{thm:invariant} shows that $d^*$ in (\ref{eq:weightederror}) is invariant under a shift of $R$ by a function of $\bm{X}$. That is, $d^*$ remains unchanged if $R$ is replaced by $R-g(\bm{X})$ for any function $g$, as long as $g$ is not related to $d$. So there is an optimal solution $\hat{f}$  in (\ref{eq:owloptimization}) corresponding to a particular function $g$, when $r_i$ is replaced by $r_i-g(\bm{x}_i)$, and the associated rule $\hat{d}_f=sign(\hat{f})$ can be seen as an estimated optimal rule in (\ref{eq:weightederror}). As $g$ varies, we obtain a collection of $\hat{d}_f$'s. When the sample size is very large, these $\hat{d}_f$'s perform similarly by the infinite sample property of OWL shown in \citet{Zhao:OWL2012}. However, when the sample size is limited, as in clinical settings, the choice of $g$ is critical to the performance of OWL. \citet{Zhao:OWL2012} did not delve deeply into this problem. In this article, we will provide a solution to this, and improve finite sample performance.
\begin{lemma} \label{thm:invariant}
For any measurable $g:\mathcal{X}\rightarrow \mathbb{R}$ and any probability distribution for $(\bm{X},A,R)$,
\begin{equation*}
\mathbb{E}\left(\frac{R-g(\bm{X})}{\pi(A,\bm{X})}\mathbb{I}\big(A\neq d(\bm{X})\big)\right)=\mathbb{E}\left(\frac{R}{\pi(A,\bm{X})}\mathbb{I}\big(A\neq d(\bm{X})\big)\right)-\mathbb{E}\left(g(\bm{X})\right).
\end{equation*}
\end{lemma}

Intuitively, the function $g$ with the smallest variance of $\displaystyle \frac{R-g(\bm{X})}{\pi(A,\bm{X})}\mathbb{I}\big(A\neq d(\bm{X})\big)$ is a good choice. However, such a function depends on the decision rule $d$, as shown in the following theorem. The proof of the theorem is provided in Appendix.
\begin{theorem} \label{thm:residual}
Among all measurable $g:\mathcal{X}\rightarrow \mathbb{R}$, $\displaystyle \tilde{g}(\bm{X})=\mathbb{E}\left(\frac{R}{\pi(A,\bm{X})}\mathbb{I}\big(A\neq d(\bm{X})\big)\Big|\bm{X}\right)$ is the function $g$ that minimizes the variance of $\displaystyle \frac{R-g(\bm{X})}{\pi(A,\bm{X})}\mathbb{I}\big(A\neq d(\bm{X})\big)$.
\end{theorem}
Our purpose is to find a function $g$, which is not related to $d$, to reduce the variance of $\displaystyle \frac{R-g(\bm{X})}{\pi(A,\bm{X})}\mathbb{I}\big(A\neq d(\bm{X})\big)$ as much as possible. As shown in the proof, the minimizer $\tilde{g}$ can be written as,
\begin{equation*}
\tilde{g}(\bm{X})=\mathbb{E}(R|\bm{X},A=1)\mathbb{I}\big(d(\bm{X})\neq1\big)+\mathbb{E}(R|\bm{X},A=-1)\mathbb{I}\big(d(\bm{X})\neq-1\big).
\end{equation*}
That is, $\tilde{g}(\bm{X})$ jumps between $\mathbb{E}(R|\bm{X},A=1)$ and $\mathbb{E}(R|\bm{X},A=-1)$ as $d(\bm{X})$ varies. Hence when $d$ is unknown, a reasonable choice of $g$ is
\begin{equation} \label{eq:gstar}
g^*(\bm{X})=\frac{\mathbb{E}(R|\bm{X},A=1)+\mathbb{E}(R|\bm{X},A=-1)}{2}=\mathbb{E}\left(\frac{R}{2\pi(A,\bm{X})}|\bm{X}\right).
\end{equation}
We propose to minimize the following empirical risk, rather than the original one in (\ref{eq:empiricleweightederror}):
\begin{equation}\label{eq:empiriclRWL}
\frac{1}{n}\sum_{i=1}^{n}\frac{r_i-\hat{g}^*(\bm{x}_i)}{\pi(a_i,\bm{x}_i)}\mathbb{I}\Big(d_f(\bm{x}_i)\neq a_i\Big),
\end{equation}
where $\hat{g}^*$ is an estimate of $g^*$. For simplicity, let $\hat{r}_i=r_i-\hat{g}^*(\bm{x}_i)$ be the estimated residual. Here, we do not weight misclassification errors by clinical outcomes as OWL does, and instead we weight them by residuals from a regression fit of outcomes. Thus the optimal decision function is invariant under any translation of clinical outcomes. We call the method Residual Weighted Learning (RWL).

\begin{figure}
\centering
\includegraphics[scale=0.8]{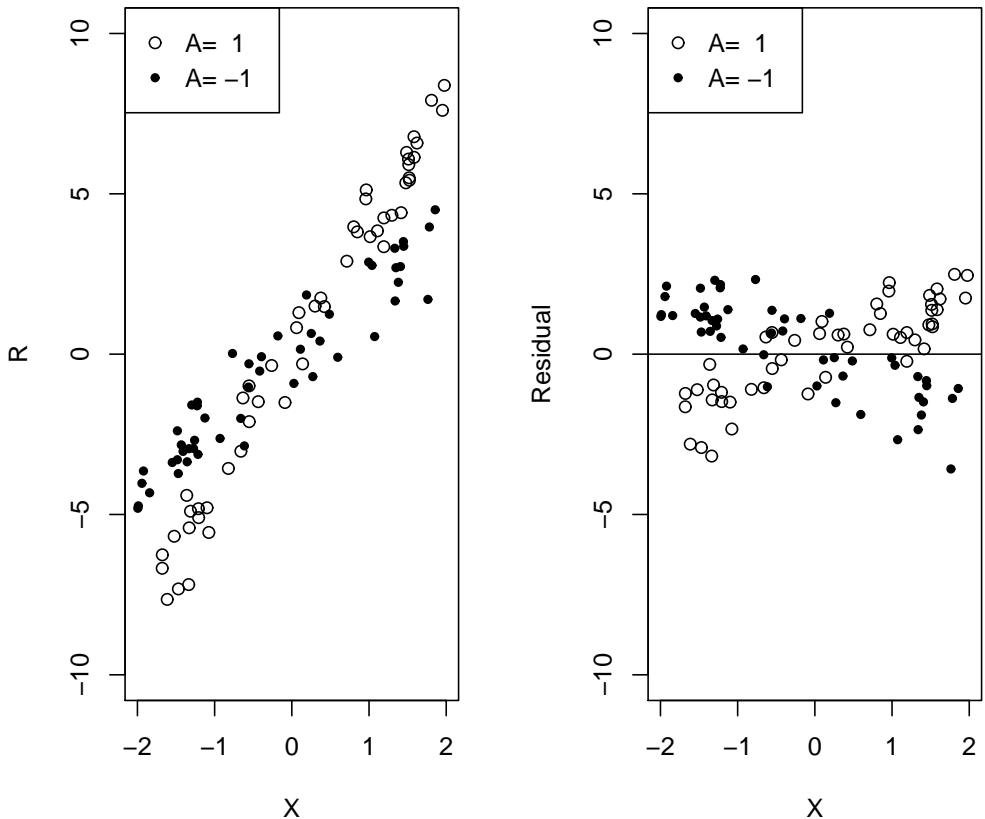}
\caption{Example of residual weighted learning. The raw data are shown on the left, consisting of a single covariate $X$, treatment assignment $A=1$ or $-1$, and continuous outcome $R$ with $E(R|X,A)=3X+XA$. The allocation ratio is 1:1. The residuals, $R-3X$, are shown on the right.}\label{fig:example}
\end{figure}

RWL also has the following interpretation. The outcome $R$ can be characterized as
\begin{equation*}
R = \mu(\bm{X})+\delta(\bm{X})\cdot A + \epsilon,
\end{equation*}
where $\epsilon$ is the mean zero random error. $\mu(\bm{X})$ reflects common effects of clinical covariates $\bm{X}$ for both treatment arms. It is easy to see that $\mu(\bm{X})=(\mathbb{E}(R|\bm{X},A=1)+\mathbb{E}(R|\bm{X},A=-1))/2$, so the residual, $R-g^*(\bm{X})$, in RWL is just $\delta(\bm{X})\cdot A+\epsilon$, which captures all sources of heterogeneous treatment effects. The rationale of optimal treatment regimes is to keep treatment assignments that subjects have actually received if those subjects are observed to have large outcomes, and to switch assignments if outcomes are small.
However, the largeness and smallness for outcomes are relative. A rather large outcome may still be considered as small when compared with subjects having similar clinical covariates, as shown in Figure~\ref{fig:example}. Outcomes are not comparable among subjects with different clinical covariates, while the residual, by removing common covariates effects, is a better measurement. Figure~\ref{fig:example} illustrates how residuals work using an example with a single covariate $X$. The raw data are shown on the left, and residuals are shown on the right. Residuals are comparable among subjects, and larger residuals represent better outcomes.

Another benefit from using residuals is a clear cut-off, \textit{i.e.} 0, to distinguish between subjects with good and poor clinical outcomes. To minimize the empirical risk in (\ref{eq:empiriclRWL}), for subjects with positive residuals, RWL is apt to recommend the same treatment assignments that subjects have actually received; for subjects with negative residuals, RWL is more likely to give the opposite treatment assignments to what they have received. The optimal ITR is the one maximizing the conditional expected outcome given in (\ref{eq:ITR}). An empirical estimator of (\ref{eq:ITR}) is,
\begin{equation} \label{eq:empiricalreward}
\displaystyle \frac{1}{n}\sum_{i=1}^{n}\frac{r_i}{\pi(a_i,\bm{x}_i)}\mathbb{I}\Big(d(\bm{x}_i)= a_i\Big).
\end{equation}
Though it is unbiased, it may give an estimate outside the range of $R$. While for finite samples a better estimator of (\ref{eq:ITR}) as shown in \citet{Murphy:Design2005} is,
\begin{equation} \label{eq:empiricalreward2}
\frac{\frac{1}{n}\sum_{i=1}^{n}\frac{r_i\mathbb{I}\big(d(\bm{x}_i)= a_i\big)}{\pi(a_i,\bm{x}_i)}}{\frac{1}{n}\sum_{i=1}^{n}\frac{\mathbb{I}\big(d(\bm{x}_i)= a_i\big)}{\pi(a_i,\bm{x}_i)}}.
\end{equation}
The denominator is an estimator of $\mathbb{E}\Big(\frac{\mathbb{I}\big(A=d(\bm{X})\big)}{\pi(A,\bm{X})}\Big)$. Similar reasoning as that used in the proof of Lemma \ref{thm:invariant} yields $\mathbb{E}\Big(\frac{\mathbb{I}\big(A=d(\bm{X})\big)}{\pi(A,\bm{X})}\Big) = 1$. The estimator $\frac{1}{n}\sum_{i=1}^{n}\frac{\mathbb{I}\big(d(\bm{x}_i)= a_i\big)}{\pi(a_i,\bm{x}_i)}$ is called the treatment matching factor in this article. The factor varies between 0 and 2. If a rule favors treatment assignments that subjects have actually received, its treatment matching factor is greater than 1, while in contrast if a rule prefers the opposite treatment assignments, its treatment matching factor is less than 1. For a randomized clinical trial, we expect that the estimated rule is associated with a treatment matching factor close to 1. For OWL, since all the weights are nonnegative, the estimated rule of OWL tends to keep, if possible, the treatments that subjects actually received. Thus the associated treatment matching factor would be greater than 1, especially when the sample size is small, or when a complicated rule is applied. Hence the estimator in (\ref{eq:empiricalreward2}) might not be large even though (\ref{eq:empiricalreward}) is maximized. RWL alleviates this problem by using residuals to make an initial guess on the optimal rule. Owing to the balance between subjects with positive and negative residuals, RWL implicitly finds a rule with its treatment matching factor close to 1.

There are many ways to estimate $g^*$. In this article, we consider two models. The first one is the main effects model. Assume that $\displaystyle \mathbb{E}\left(\frac{R}{2\pi(A,\bm{X})}|\bm{X}\right)=\beta_0+\bm{X}^T\bm{\beta}$, where $\bm{\beta}=(\beta_1,\cdots,\beta_p)^T$. The estimates $\hat{\bm{\beta}}$ and $\hat{\beta}_0$ can be obtained by minimizing the sum of weighted squares,
\begin{equation*}
\sum_{i=1}^n\frac{1}{2\pi(a_i,\bm{x}_i)}(r_i-\beta_0-\bm{x}_i^T\bm{\beta})^2.
\end{equation*}
It can be solved easily by almost any statistical software. Then the estimate is $\hat{g}^*(\bm{x})=\hat{\beta}_0+\bm{x}^T\bm{\hat{\beta}}$. The second is the null model. Assume that $\displaystyle \mathbb{E}\left(\frac{R}{2\pi(A,\bm{X})}|\bm{X}\right)=\beta_0$. It is easy to obtain $\hat{g}^*(\bm{x}) = \hat{\beta}_0 = \sum_{i=1}^n\frac{r_i}{\pi(a_i,\bm{x}_i)}/\sum_{i=1}^n\frac{1}{\pi(a_i,\bm{x}_i)}$.

In summary, we propose a method called Residual Weighted Learning to identify the optimal ITR by minimizing the residual weighted classification error (\ref{eq:empiriclRWL}). The impact of using residuals is two-fold: it stabilizes the variance of the value function and controls the treatment matching factor. Both improve finite sample performance.

\subsection{Implementation of RWL}
\begin{figure}
\centering
\subfigure[Hinge loss]{\label{fig:hingeloss} \includegraphics[scale=0.5]{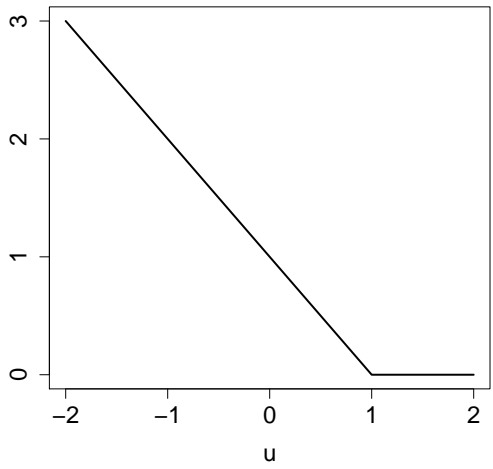}} 
\subfigure[Ramp loss]{\label{fig:ramploss}
\includegraphics[scale=0.5]{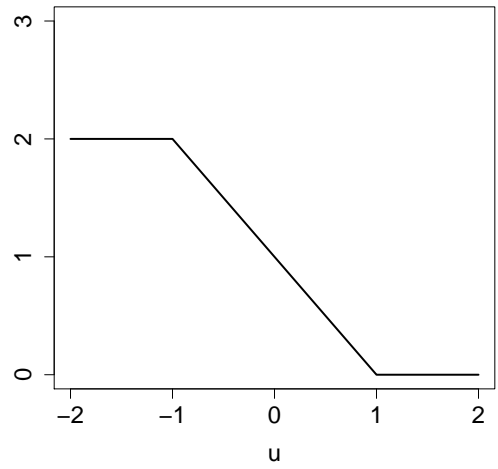}} \\

\subfigure[Smoothed ramp loss]{\label{fig:smoothedramploss} \includegraphics[scale=0.5]{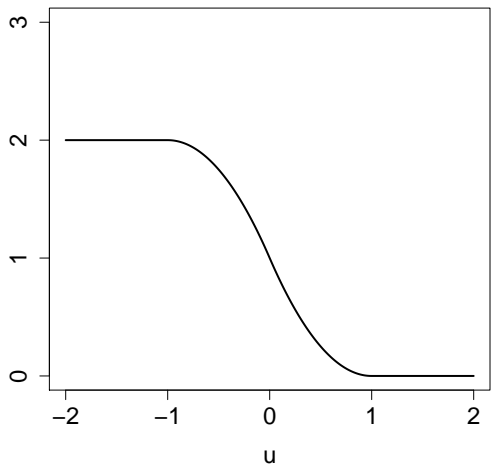}} 
\subfigure[Decomposition]{\label{fig:ccdecomp}
\includegraphics[scale=0.5]{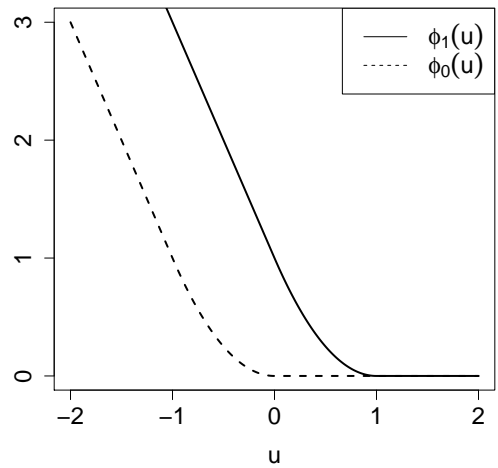}}
\caption{Hinge loss (a), ramp loss (b), and smoothed ramp loss (c) functions. (d) shows the difference of convex decomposition of the smoothed ramp loss, $T(u) = \phi_1(u) - \phi_0(u)$.}
\label{fig:lossfunction}
\end{figure}

As in OWL, we intend to use a surrogate loss function to replace the 0-1 loss in (\ref{eq:empiriclRWL}). Since some residuals are negative, convex surrogate functions do not work here. We consider a non-convex loss
\begin{equation*}
T(u) = \left\{
\begin{array}{ll}
0 & {\rm if}\: u\geq1, \\
(1-u)^2 & {\rm if}\: 0\leq u<1, \\
2-(1+u)^2 & {\rm if}\: -1\leq u<0, \\
2 & {\rm if}\: u<-1. \\
\end{array}
\right.
\end{equation*}
It is called the smoothed ramp loss in this article. Figure~\ref{fig:lossfunction} shows the hinge loss, ramp loss and smoothed ramp loss functions. The hinge loss is the loss function used in support vector machines \citep{Vapnik:svm98}. The ramp loss \citep{Collobert2006:RampLoss} is also called the truncated hinge loss function \citep{Wu2007:RobustSVM}. It is well known that by truncating the unbounded hinge loss, ramp loss related methods are shown to be robust to outliers in the training data for the classification problem \citep{Wu2007:RobustSVM}. Compared with the ramp loss, the smoothed ramp loss is smooth everywhere. Hence it has computational advantages in optimization. Moreover, the smoothed ramp loss, which resembles the ramp loss, is robust to outliers too.
In the framework of RWL, subjects who did not receive optimal treatment assignments but had large positive residuals, or those who did receive optimal assignments but had large negative residuals may be considered as outliers. The robustness to outliers for the smoothed ramp loss is helpful in dealing with outliers in the setting of optimal treatment regimes, especially when residuals are poorly estimated, or when the outcome $R$ has a large variance.

We incorporate RWL into the regularization framework, and aim to minimize
\begin{equation} \label{eq:rwlramploss}
\frac{1}{n}\sum_{i=1}^{n}\frac{\hat{r}_i}{\pi(a_i,\bm{x}_i)}T(a_if(\bm{x}_i))+\frac{\lambda}{2}||f||^2,
\end{equation}
where $||f||$ is some norm for $f$, and $\lambda$ is a tuning parameter. Recall that $\hat{r}_i$ is the estimated residual.
The smoothed ramp loss function $T(u)$ is symmetric about the point $(0,1)$ as shown in Figure~\ref{fig:smoothedramploss}. A nice property that comes from the symmetry is that the rule that minimizes the outcome $R$ (\textit{i.e.} maximizes $-R$) is just opposite to the rule that maximizes $R$. This is intuitively sensible. However, OWL does not possess this property.

We derive an algorithm for linear RWL in Section~\ref{sec:linearrwl}, and then generalize it to the case of nonlinear learning through kernel mapping in Section~\ref{sec:nonlinearrwl}.

\subsubsection{Linear Decision Rule for Optimal ITR} \label{sec:linearrwl}
Suppose that the decision function $f(\bm{x})$ that minimizes (\ref{eq:rwlramploss}) is a linear function of $\bm{x}$, \textit{i.e.} $f(\bm{x})=\bm{w}^T\bm{x}+b$. Then the associated ITR will assign a subject with clinical covariates $\bm{x}$ into treatment 1 if $\bm{w}^T\bm{x}+b>0$ and $-1$ otherwise. In (\ref{eq:rwlramploss}), we define $||f||$ as the Euclidean norm of $\bm{w}$. Then minimizing (\ref{eq:rwlramploss}) can be rewritten as
\begin{equation}
\min_{\bm{w},b} \quad \frac{\lambda}{2}\bm{w}^T\bm{w} + \frac{1}{n}\sum_{i=1}^{n}\frac{\hat{r}_i}{\pi(a_i,\bm{x}_i)}T\big(a_i(\bm{w}^T\bm{x}_i+b)\big).
\label{eq:linearrwl}
\end{equation}
The smoothed ramp loss is a non-convex loss, and as a result, the optimization problem in (\ref{eq:linearrwl}) involves non-convex minimization. This optimization problem is difficult since there are many local minima or stationary points. For instance, any $(\bm{w},b)$ with $\bm{w}=\bm{0}$ and $|b|\geq 1$ is a stationary point.
Similar with the robust truncated hinge loss support vector machine \citep{Wu2007:RobustSVM}, we apply the d.c. (Difference of Convex) algorithm \citep{An1997:DC} to solve this non-convex minimization problem. The d.c. algorithm is also known as the Concave-Convex Procedure (CCCP) in the machine learning community \citep{Yuille2003:CCCP}. Assume that an objective function can be rewritten as the sum of a convex part $Q_{vex}(\Theta)$ and a concave part $Q_{cav}(\Theta)$. The d.c. algorithm as shown in Algorithm 1 solves the non-convex optimization problem by minimizing a sequence of convex subproblems.
One can easily see that the d.c. algorithm is a special case of the Majorize-Minimization (MM) algorithm. 
\begin{algorithm} \label{alg:DC}
  \DontPrintSemicolon
  \caption{The d.c. algorithm for minimizing $Q(\Theta)=Q_{vex}(\Theta)+Q_{cav}(\Theta)$.}
  Initialize $\Theta^{(0)}$ \;
  \Repeat{convergence of $\Theta^{(t)}$}{
    $\Theta^{(t+1)}=\textrm{argmin}_{\Theta} Q_{vex}(\Theta)+<Q'_{cav}(\Theta^{(t)}),\Theta>$\;
  }
\end{algorithm}

Let
\begin{equation*}
\phi_{s}(u) = \left\{
\begin{array}{ll}
0 & {\rm if}\: u\geq s \\
(s-u)^2 & {\rm if}\: s-1\leq u<s \\
2s-2u-1 & {\rm if}\: u< s-1 \\
\end{array}
\right..
\end{equation*}
Note that $\phi_{s}$ is smooth. We have a difference-of-convex decomposition of the smoothed ramp loss,
\begin{equation} \label{eq:sramploss_decomp}
T(u) = \phi_1(u) - \phi_0(u),
\end{equation}
as shown in Figure~\ref{fig:ccdecomp}. Denote $\Theta$ as $(\bm{w},b)$. Applying (\ref{eq:sramploss_decomp}), the objective function in (\ref{eq:linearrwl}) can be decomposed as
\begin{eqnarray*}
Q^s(\Theta)& = & \underbrace{\frac{\lambda}{2}\bm{w}^T\bm{w}+\frac{1}{n}\sum_{i=1}^n\Big[\phi_1(u_i)\mathbb{I}(\hat{r}_i>0)+\phi_{0}(u_i)\mathbb{I}(\hat{r}_i<0)\Big]\frac{|\hat{r}_i|}{\pi(a_i,\bm{x}_i)}}_{Q_{vex}(\Theta)}
\\
& & \underbrace{-\frac{1}{n}\sum_{i=1}^n\Big[\phi_1(u_i)\mathbb{I}(\hat{r}_i<0)+\phi_{0}(u_i)\mathbb{I}(\hat{r}_i>0)\Big]\frac{|\hat{r}_i|}{\pi(a_i,\bm{x}_i)}}_{Q_{cav}(\Theta)},
\end{eqnarray*}
where $u_i = a_i(\bm{w}^T\bm{x}_i+b)$.
For simplicity, we introduce the notation,
\begin{equation} \label{eq:beta}
\beta_i = \frac{\partial Q_{cav}}{\partial u_i}=-\frac{1}{n}\Big[\frac{d \phi_1(u_i)}{d u_i}\mathbb{I}(\hat{r}_i<0)+\frac{d \phi_0(u_i)}{d u_i}\mathbb{I}(\hat{r}_i>0)\Big]\frac{|\hat{r}_i|}{\pi(a_i,\bm{x}_i)},
\end{equation}
for $i=1,\cdots,n$. Thus the convex subproblem at the $(t+1)$'th iteration of the d.c. algorithm is
\begin{equation}
\min_{\bm{w},b} \quad \frac{\lambda}{2}\bm{w}^T\bm{w} + \frac{1}{n}\sum_{i=1}^n\Big[\phi_1(u_i)\mathbb{I}(\hat{r}_i>0)+\phi_{0}(u_i)\mathbb{I}(\hat{r}_i<0)\Big]\frac{|\hat{r}_i|}{\pi(a_i,\bm{x}_i)}+\sum_{i=1}^{n}\beta_i^{(t)}u_i.
\label{eq:subproblem}
\end{equation}
There are many efficient methods available for solving smooth unconstrained optimization problems. In this article, we use the limited-memory Broyden-Fletcher-Goldfarb-Shanno (L-BFGS) algorithm \citep{Nocedal1980:LBFGS}. L-BFGS is a quasi-Newton method that approximates the Broyden-Fletcher-Goldfarb-Shanno (BFGS) algorithm using a limited amount of computer memory. For more information on the quasi-Newton method and L-BFGS, see \citet{Nocedal2006:NO}.

The procedure is summarized in the following algorithm:
\begin{algorithm} \label{alg:RWLLinear}
  \DontPrintSemicolon
  \caption{The d.c. algorithm for linear RWL with the smoothed ramp loss}
  Set $\epsilon$ to a small quantity, say, $10^{-8}$ \;
  Initialize $\bm{\beta}_i^{(0)}=\frac{2|\hat{r}_i|}{n\pi(a_i,\bm{x}_i)}\mathbb{I}(\hat{r}_i<0)$ \;
  \Repeat{$||\bm{\beta}^{(t+1)}-\bm{\beta}^{(t)}||_{\infty}<\epsilon$}{
  \begin{itemize}
  \item Compute $(\hat{\bm{w}}, \hat{b})$ by solving (\ref{eq:subproblem}),
  \item Update $u_i = a_i(\hat{\bm{w}}^T\bm{x}_i+\hat{b})$
  \item Update $\beta_i^{(t+1)}$ by (\ref{eq:beta})
  \end{itemize}
  }
\end{algorithm}

\subsubsection{Nonlinear Decision rule for Optimal ITR} \label{sec:nonlinearrwl}
For a nonlinear decision rule, the decision function $f(\bm{x})$ is represented by $h(\bm{x})+b$ with $h(\bm{x})\in \mathcal{H}_K$ and $b\in\mathbb{R}$, where $\mathcal{H}_K$ is a reproducing kernel Hilbert space (RKHS) associated with a Mercer kernel function $K$. The kernel function $K(\cdot,\cdot)$ is a positive definite function mapping from $\mathcal{X}\times\mathcal{X}$ to $\mathbb{R}$. The norm in $\mathcal{H}_K$, denoted by $||\cdot||_K$, is induced by the following inner product:
\begin{equation*}
<f,g>_K=\sum_{i=1}^{n}\sum_{j=1}^{m}\alpha_i\beta_jK(\bm{x}_i,\bm{x}_j),
\end{equation*}
for $f(\cdot)=\sum_{i=1}^n\alpha_iK(\cdot,\bm{x}_i)$ and $g(\cdot)=\sum_{j=1}^m\beta_jK(\cdot,\bm{x}_j)$. Then minimizing (\ref{eq:rwlramploss}) can be rewritten as
\begin{equation}
\min_{h,b} \quad \frac{\lambda}{2}||h||_K^2 + \frac{1}{n}\sum_{i=1}^{n}\frac{\hat{r}_i}{\pi(a_i,\bm{x}_i)}T\Big(a_i\big(h(\bm{x}_i)+b\big)\Big).
\label{eq:nonlinearrwl}
\end{equation}
Due to the representer theorem  \citep{Kimeldorf:Representer}, the nonlinear problem can be reduced to finding finite-dimensional coefficients $v_i$, and $h(\bm{x})$ can be represented as $\sum_{j=1}^nv_jK(\bm{x},\bm{x}_j)$. Thus the problem~(\ref{eq:nonlinearrwl}) is transformed to
\begin{equation} \label{nonlinearrwlprime}
\min_{\bm{v},b} \quad \frac{\lambda}{2}\sum_{i,j=1}^nv_iv_jK(\bm{x}_i,\bm{x}_j) + \frac{1}{n}\sum_{i=1}^{n}\frac{\hat{r}_i}{\pi(a_i,\bm{x}_i)}T\Big(a_i\big(\sum_{j=1}^nv_jK(\bm{x}_i,\bm{x}_j)+b\big)\Big).
\end{equation}
Following a similar derivation to that used in the previous section, the convex subproblem at the $(t+1)$'th iteration of the d.c. algorithm is as follows,
\begin{equation*}
\min_{\bm{v},b} \quad \frac{\lambda}{2}\sum_{i,j=1}^nv_iv_jK(\bm{x}_i,\bm{x}_j) + \frac{1}{n}\sum_{i=1}^n\Big[\phi_1(u_i)\mathbb{I}(\hat{r}_i>0)+\phi_{0}(u_i)\mathbb{I}(\hat{r}_i<0)\Big]\frac{|\hat{r}_i|}{\pi(a_i,\bm{x}_i)}+\sum_{i=1}^{n}\beta_i^{(t)}u_i,
\end{equation*}
where $u_i = a_i\big(\sum_{j=1}^nv_jK(\bm{x}_i,\bm{x}_j)+b\big)$.
After solving the subproblem by L-BFGS, we update $\beta^{(t+1)}$ by (\ref{eq:beta}). The procedure is repeated until $\bm{\beta}$ converges. When we obtain the solution $(\hat{\bm{v}},\hat{b})$, the decision function is $\hat{f}(\bm{x})=\sum_{j=1}^n\hat{v}_jK(\bm{x},\bm{x}_j)+\hat{b}$. Note that if we choose a linear kernel $K(\bm{x},\bm{z})=\bm{x}^T\bm{z}$, the obtained rule reduces to the previous linear rule. The most widely used nonlinear kernel in practice is the Gaussian Radial Basis Function (RBF) kernel, that is,
\begin{equation*}
K_{\sigma}(\bm{x},\bm{z})=\exp\Big(-\sigma^2||\bm{x}-\bm{z}||^2\Big),
\end{equation*}
where $\sigma>0$ is a free parameter whose inverse $1/\sigma$ is called the width of $K_{\sigma}$.

\subsection{A general framework for Residual Weighted Learning} \label{sec:generalframework}
We have proposed a method to identify the optimal ITR for continuous outcomes. However, in clinical practice, the endpoint outcome could also be binary, count, or rate. In this section, we provide a general framework to deal with all these types of outcomes.

The procedure is described as follows. First, estimate the conditional expected outcomes given clinical covariates $\bm{X}$, $\hat{\mathbb{E}}\left(\frac{R}{2\pi(A,\bm{X})}\big|\bm{X}\right)$, by an appropriate regression model. It is equivalent to fitting a weighted regression model, in which each subject is weighted by $\frac{1}{2\pi(A,\bm{X})}$. Second, we calculate the estimated residual $\hat{r}_i$ by comparing the observed outcome and expected outcome estimated in the first step. If the observed outcome is better than the expected one, the estimated residual $\hat{r}_i$ is positive, otherwise it is negative. Specifically, if larger values of $R$ are more desirable, as with our assumption for continuous outcomes, $\hat{r}_i = r_i - \hat{\mathbb{E}}\left(\frac{R}{2\pi(A,\bm{X})}\big|\bm{X}=\bm{x}_i\right)$; if smaller values of $R$ are preferred, \textit{e.g.} the number of adverse events,  then $\hat{r}_i = \hat{\mathbb{E}}\left(\frac{R}{2\pi(A,\bm{X})}\big|\bm{X}=\bm{x}_i\right)-r_i$.
Third, identify the decision function $\hat{f}(\bm{x})$ by using the estimated $\hat{r}_i$ in RWL.

The underlying idea is simple. Generally, the outcome $R$ is a random variable depending on $\bm{X}$ and $A$. We estimate common effects of $\bm{X}$ by ignoring the treatment assignment $A$, and then all information on the heterogeneous treatment effects is contained in the residuals. As demonstrated previously, the benefits from using residuals include stabilizing the variance and controlling the treatment matching factor, both of which have a positive impact on finite sample performance. In previous sections, we discussed RWL for continuous outcomes in detail. We will provide two additional examples to illustrate the general framework.

The first example is for binary outcomes, \textit{i.e.} $R\in\{0,1\}$. Assume that $R=1$ is desirable. We may fit a weighted main effects logistic regression model,
\begin{equation*}
\mathbb{E}(R|\bm{X}) = \frac{\exp(\beta_0+\bm{X}^T\bm{\beta})}{1+\exp(\beta_0+\bm{X}^T\bm{\beta})}.
\end{equation*}
The estimates $\hat{\beta}_0$ and $\hat{\bm{\beta}}$ can be obtained numerically by statistical software, for example, the glm function in R. The residual $\hat{r}_i$ is $r_i-\frac{\exp(\hat{\beta}_0+\bm{X}^T\hat{\bm{\beta}})}{1+\exp(\hat{\beta}_0+\bm{X}^T\hat{\bm{\beta}})}$. Then we estimate the optimal ITR by RWL.

The second is for count/rate outcomes. We use the pulmonary exacerbation (PE) outcome of the cystic fibrosis data in Section \ref{sec:dataanalysis} as an example. The outcome is the number of PEs during the study, and it is a rate variable. The observed data are $D_n=\{\bm{x}_i, a_i, r_i, t_i\}_{i=1}^n$, where $r_i$ is the number of PEs in the duration $t_i$. We treat the count $r_i$ as the outcome, and $(\bm{x}_i, t_i)$ as clinical covariates. We may fit a weighted main effects Poisson regression model,
\begin{equation*}
\log\left(\mathbb{E}(R|\bm{X},T)\right) = \beta_0+\bm{X}^T\bm{\beta}+\log(T).
\end{equation*}
Since fewer PEs are desirable, we compute $\hat{r}_i$ as $\exp\left(\hat{\beta}_0+\bm{x}_i^T\hat{\bm{\beta}}+\log(t_i)\right)-r_i$, which is the opposite of the raw residual in generalized linear models.

\section{Theoretical Properties}
In this section, we establish theoretical results for RWL. To the end, we assume that a sample $D_n=\{\bm{x}_i, a_i, r_i\}_{i=1}^n$, is independently drawn from a probability measure $P$ on $\mathcal{X}\times\mathcal{A}\times\mathcal{M}$, where $\mathcal{X}\subset \mathbb{R}^p$ is compact, and $\mathcal{M}=[-M,M]\subset\mathbb{R}$. For any ITR $d: \mathcal{X}\rightarrow \mathcal{A}$, the risk is defined as
\begin{equation*}
\mathcal{R}(d) = \mathbb{E}\left[\frac{R}{\pi(A,\bm{X})}\mathbb{I}(A\neq d(\bm{X}))\right].
\end{equation*}
The ITR that minimizes the risk is the Bayes rule $d^*=\arg\min_{d}\mathcal{R}(d)$, and the corresponding risk $\mathcal{R}^*=\mathcal{R}(d^*)$ is the Bayes risk. Recall that the Bayes rule is $d^*(\bm{x})=sign(\mathbb{E}(R|\bm{X}=\bm{x},A=1)-\mathbb{E}(R|\bm{X}=\bm{x},A=-1))$.

In RWL, we substitute the 0-1 loss $\mathbb{I}(A\neq sign(f(\bm{X})))$ by the smoothed ramp loss, $T(Af(\bm{X}))$. Accordingly, we define the $T$-risk as
\begin{equation*}
\mathcal{R}_{T,g}(f) = \mathbb{E}\left[\frac{R-g(\bm{X})}{\pi(A,\bm{X})}T(Af(\bm{X}))\right],
\end{equation*}
and, similarly, the minimal $T$-risk as $\mathcal{R}_{T,g}^*=\inf_f \mathcal{R}_{T,g}(f)$ and $f_{T,g}^*=\arg\min_f \mathcal{R}_{T,g}(f)$. In the theoretical analysis, we do not require $g$ to be a regression fit of $R$. $g$ can be any arbitrary measurable function. We may further assume that $(R-g(\bm{X}))/\pi(A,\bm{X})$ is bounded almost surely.

The performance of an ITR associated with a real-valued function $f$ is measured by the excess risk $\mathcal{R}(sign(f))-\mathcal{R}^*$. In terms of the value function, we note that the excess risk is just  $\mathcal{V}(d^*)-\mathcal{V}(sign(f))$. Similarly, we define the excess $T$-risk as $\mathcal{R}_{T,g}(f)-\mathcal{R}_{T,g}^*$.

Let $f_{D_n,\lambda_n}\in \mathcal{H}_K+\{1\}$, \textit{i.e.} $f_{D_n,\lambda_n}=h_{D_n,\lambda_n}+b_{D_n,\lambda_n}$ where $h_{D_n,\lambda_n}\in \mathcal{H}_K$ and $b_{D_n,\lambda_n}\in \mathbb{R}$, be a global minimizer of the following optimization problem:
\begin{equation*}
\min_{f=h+b\in\mathcal{H}_K+\{1\}} \quad \frac{\lambda_n}{2}||h||_K^2 + \frac{1}{n}\sum_{i=1}^{n}\frac{r_i-g(\bm{x}_i)}{\pi(a_i,\bm{x}_i)}T\Big(a_if(\bm{x}_i))\Big).
\end{equation*}
Here we suppress $g$ from the notation of $f_{D_n,\lambda_n}$, $h_{D_n,\lambda_n}$ and $b_{D_n,\lambda_n}$.

The purpose of the theoretical analysis is to estimate the excess risk $\mathcal{R}(sign(f_{D_n,\lambda_n}))-\mathcal{R}^*$ as the sample size $n$ tends to infinity. Convergence rates will be derived under the choice of the parameter $\lambda_n$ and conditions on the distribution $P$.

\subsection{Fisher Consistency}
We establish Fisher consistency of the decision function based on the smoothed ramp loss. Specifically, the following result holds:
\begin{theorem} \label{thm:fisher}
For any measurable function $f$, if ${f}^*_{T,g}$ minimizes T-risk $\mathcal{R}_{T,g}(f)$, then the Bayes rule $d^*(\bm{x})=sign({f}^*_{T,g}(\bm{x}))$ for all $\bm{x}$ such that $\mathbb{E}(R|\bm{X}=\bm{x},A=1) \neq \mathbb{E}(R|\bm{X}=\bm{x},A=-1)$. Furthermore, $\mathcal{R}_{T,g}(d^*)=\mathcal{R}_{T,g}^*$.
\end{theorem}
The proof is provided in Appendix. This theorem shows the validity of using the smoothed ramp loss as a surrogate loss in RWL.

\subsection{Excess Risk}
The following result establishes the relationship between the excess risk and excess $T$-risk. The proof can be found in Appendix.
\begin{theorem} \label{thm:excessrisk}
For any measurable $f:\mathcal{X}\rightarrow \mathbb{R}$ and any probability distribution for $(\bm{X},A,R)$,
\begin{equation*}
\mathcal{R}(sign(f))-\mathcal{R}^*\leq \mathcal{R}_{T,g}(f)-\mathcal{R}_{T,g}^*.
\end{equation*}
\end{theorem}
This theorem shows that for any decision function $f$, the excess risk of $f$ under the 0-1 loss is no larger than the excess risk of $f$ under the smoothed ramp loss. It implies that if we obtain $f$ by approximately minimizing $\mathcal{R}_{T,g}(f)$ so that the excess $T$-risk is small, then the risk of $f$ is close to the Bayes risk. In the next two sections, we investigate properties of $f_{D_n,\lambda_n}$ through the excess $T$-risk.

\subsection{Universal Consistency}

In the literature of machine learning, a classification rule is called universally consistent if its expected error probability converges in probability to the Bayes error probability for all distributions underlying the data \citep{Devroye:PatternRecog1996}. We extend this concept to ITRs. Precisely, given a sequence $D_n$ of training data, an ITR $d_n$ is said to be \textit{universally consistent} if
\begin{equation*}
\lim_{n\rightarrow\infty}\mathcal{R}(d_n) = \mathcal{R}^*
\end{equation*}
holds in probability for all probability measures $P$ on $\mathcal{X}\times\mathcal{A}\times\mathcal{M}$. In this section, we will establish universal consistency of the rule $d_{D_n,\lambda_n}=sign(f_{D_n,\lambda_n})$. Before proceeding to the main result, we introduce some preliminary background on RKHSs, which is essential for our work.

Let $K$ be a kernel, and $\mathcal{H}_K$ be its associated RKHS. We often use the quantity
\begin{equation*}
C_K=\sup_{\bm{x}\in\mathcal{X}}\sqrt{K(\bm{x},\bm{x})}.
\end{equation*}
In this article, we assume $C_K$ is finite. For the Gaussian RBF kernel, $C_K=1$. The assumption also holds for the linear kernel when $\mathcal{X}\subset \mathbb{R}^p$ is compact. By the reproducing property, it is easy to see that $||f||_{\infty}\leq C_K||f||_K$ for any $f\in\mathcal{H}_K$. A continuous kernel $K$ on a compact metric space $\mathcal{X}$ is called \textit{universal} if its associated RKHS $\mathcal{H}_K$ is dense in $C(\mathcal{X})$, \textit{i.e.}, for every function $g\in C(\mathcal{X})$ and all $\epsilon>0$ there exists an $f\in\mathcal{H}_K$ such that $||f-g||_{\infty}<\epsilon$ \citep[Definition 4.52]{Steinwart:SVM2008}. $C(\mathcal{X})$ is the space of all continuous functions $f:\mathcal{X}\rightarrow\mathbb{R}$ on the compact metric space $\mathcal{X}$ endowed with the usual supremum norm. The widely used Gaussian RBF kernel is an example of a universal kernel.


We are ready to present the main result of this section. The following theorem shows the convergence of the $T$-risk on the sample dependent function $f_{D_n,\lambda_n}$. We apply empirical process techniques to show consistency. The proof is provided in Appendix.
\begin{theorem} \label{thm:consistency}
Assume that we choose a sequence $\lambda_n>0$ such that $\lambda_n\rightarrow0$ and $n\lambda_n\rightarrow\infty$. For a measurable function $g$, assume that $\frac{|R-g(\bm{X})|}{\pi(A,\bm{X})}\leq M_0$ almost surely. Then for any distribution $P$ for $(\bm{X},A,R)$, we have that in probability,
\begin{equation*}
\lim_{n\rightarrow\infty}\mathcal{R}_{T,g}(f_{D_n,\lambda_n}) = \inf_{f\in\mathcal{H}_K+\{1\}}\mathcal{R}_{T,g}(f).
\end{equation*}
\end{theorem}

By Theorem~\ref{thm:fisher}, the Bayes rule $d^*$ minimizes the $T$-risk $\mathcal{R}_{T,g}(f)$. Universal consistency follows if $\inf_{f\in\mathcal{H}_K+\{1\}}\mathcal{R}_{T,g}(f)=\mathcal{R}_{T,g}^*$ by Theorem \ref{thm:excessrisk}. That is, $d^*$ needs to be approximated by functions in the space ${H}_K+\{1\}$. Clearly, this is not always true. A counterexample is when $K$ is linear.

Let $\mu$ be the marginal distribution of $\bm{X}$. Clearly, the rule $d^*$ is measurable with respect to $\mu$. Recall that a probability measure is regular if it is defined on the Borel sets. By Lusin's theorem, a measurable function can be approximated by a continuous function when $\mu$ is regular \citep[Theorem 2.24]{Rudin:Measure1987}. Thus the RKHS $\mathcal{H}_K$ of a universal kernel $K$ is rich enough to provide arbitrarily accurate decision rules for all distributions. The finding is summarized in the following universal approximation lemma. The rigorous proof is provided in Appendix.
\begin{lemma} \label{thm:universalapprox}
Let $K$ be a universal kernel, and $\mathcal{H}_K$ be the associated RKHS. For all distributions $P$ with regular marginal distribution $\mu$ on $\bm{X}$, we have
\begin{equation*}
\inf_{f\in\mathcal{H}_K+\{1\}}\mathcal{R}_{T,g}(f) = \mathcal{R}_{T,g}^*.
\end{equation*}
\end{lemma}

This immediately leads to universal consistency of RWL with a universal kernel.
\begin{proposition}
Let $K$ be a universal kernel, and $\mathcal{H}_K$ be the associated RKHS. Assume that we choose a sequence $\lambda_n>0$ such that $\lambda_n\rightarrow0$ and $n\lambda_n\rightarrow\infty$. For a measurable function $g$, assume that $\frac{|R-g(\bm{X})|}{\pi(A,\bm{X})}\leq M_0$ almost surely. Then for any distribution $P$ for $(\bm{X},A,R)$ with regular marginal distribution on $\bm{X}$, we have that in probability,
\begin{equation*}
\lim_{n\rightarrow\infty}\mathcal{R}(sign(f_{D_n,\lambda_n})) = \mathcal{R}^*.
\end{equation*}
\end{proposition}

\subsection{Convergence Rate}
A consistent rule guarantees that by increasing the amount of data the rule can eventually learn the optimal decision with high probability. The next natural question is whether there are rules $d_n$ with $\mathcal{R}(d_n)$ tending to the Bayes risk $\mathcal{R}^*$ at a specified rate for all distributions. Unfortunately, this is impossible as shown in the following theorem. The proof follows the arguments in \citet[Theorem 7.2]{Devroye:PatternRecog1996}, which states in the classification problem the rate of convergence of any particular rule to the Bayes risk may be arbitrarily slow. We provide the outline of the proof in Appendix.
\begin{theorem} \label{thm:slowrate}
Assume there are infinite distinct points in $\mathcal{X}$. Let $\{c_n\}$ be a sequence of positive numbers converging to zero with $\frac{1}{16}\geq c_1 \geq c_2\geq \cdots$. For every sequence $d_n$ of ITRs, there exists a distribution of $(\bm{X},A,R)$ on $\mathcal{X}\times\mathcal{A}\times\mathcal{M}$ such that
\begin{equation*}
\mathcal{R}(d_n) - \mathcal{R}^* \geq 2Mc_n.
\end{equation*}
\end{theorem}

This is a generic result, and applies to any algorithm for ITRs. It implies that rate of convergence studies for particular rules must necessarily be accompanied by conditions on $(\bm{X},A,R)$. Before moving back to RWL, we introduce the following quantity:
\begin{equation*}
\mathcal{A}(\lambda) = \inf_{h\in \mathcal{H}_K, b\in\mathbb{R}}\frac{\lambda}{2}||h||_K^2+\mathcal{R}_{T,g}(h+b) - \mathcal{R}^*_{T,g}.
\end{equation*}
The term $\mathcal{A}(\lambda)$ describes how well the infinite-sample regularized $T$-risk $\frac{\lambda}{2}||h||_K^2+\mathcal{R}_{T,g}(h+b)$ approximates the optimal $T$-risk $\mathcal{R}^*_{T,g}$. A similar quantity is called the \textit{approximation error function} in the literature of learning theory \citep[Definition 5.14]{Steinwart:SVM2008}. Since $\mathcal{A}(\lambda)$ is an infimum of affine linear functions, $\mathcal{A}(\lambda)$ is continuous. Thus
\begin{equation*}
\lim_{\lambda\rightarrow0}\mathcal{A}(\lambda)  = \inf_{h\in \mathcal{H}_K, b\in\mathbb{R}}\mathcal{R}_{T,g}(h+b) - \mathcal{R}^*_{T,g}.
\end{equation*}
By Lemma \ref{thm:universalapprox}, for a universal kernel $K$, $\lim_{\lambda\rightarrow0}\mathcal{A}(\lambda)=0$.

In order to establish the convergence rate, let us additionally assume that there exist constants $c>0$ and $\beta\in(0,1]$ such that
\begin{equation} \label{approximateassumption}
\mathcal{A}(\lambda)\leq c\lambda^{\beta}, \quad \forall \lambda>0.
\end{equation}
The assumption is standard in the literature of learning theory \citep[p. 229]{Steinwart:SVM2008}. The following theorem is the main result in this section. This theorem is proved in Appendix using techniques of concentration inequalities developed in \citet{Bartlett:Rademacher2002}.

\begin{theorem} \label{thm:convergencerate}
For any distribution $P$ for $(\bm{X},A,R)$ that satisfies the approximation error assumption (\ref{approximateassumption}), take $\lambda_n=n^{-\frac{1}{2\beta+1}}$. For a measurable function $g$, assume that $\frac{|R-g(\bm{X})|}{\pi(A,\bm{X})}\leq M_0$ almost surely.
Then with probability at least $1-\delta$,
\begin{equation*}
\mathcal{R}(sign(f_{D_n,\lambda_n})) - \mathcal{R}^* \leq \tilde{c}\sqrt{\log(4/\delta)}n^{-\frac{\beta}{2\beta+1}},
\end{equation*}
where the constant $\tilde{c}$ is independent of $\delta$ and $n$, and $\tilde{c}$ decreases as $M_0$ decreases.
\end{theorem}

The resulting rate depends on the polynomial decay rate $\beta$ of the approximation error, and the optimal rate is about $n^{-1/3}$  when $\beta$ approaches 1. Theorem \ref{thm:convergencerate} also implies that even when the sample size $n$ is fixed, an appropriately chosen $g$ with a small bound $M_0$ would improve performance. It is easy to verify that $g^*$ in (\ref{eq:gstar}) is the function $g$ that minimizes $\mathbb{E}\left(\frac{(R-g(\bm{X}))^2}{\pi(A,\bm{X})}\right)$. So when we choose $g^*$ as in RWL, we may possibly obtain a smaller $M_0$, as shown in Figure~\ref{fig:example}.

We are particularly interested in the Gaussian RBF kernel since it is widely used and is universal. It is important to understand the approximation error assumption (\ref{approximateassumption}) for the Gaussian RBF kernel. We first introduce the following ``geometric noise'' assumption for $P$ on $(\bm{X},A,R)$ \citep{Steinwart:FastRateGaussian}. Let
\begin{equation*}
\delta(\bm{x})=\mathbb{E}(R|\bm{X}=\bm{x},A=1) - \mathbb{E}(R|\bm{X}=\bm{x},A=-1).
\end{equation*}
Define $\mathcal{X}^+=\{\bm{x}\in\mathcal{X}: \delta(\bm{x})>0\}$,  $\mathcal{X}^-=\{\bm{x}\in\mathcal{X}: \delta(\bm{x})<0\}$, and $\mathcal{X}^0=\{\bm{x}\in\mathcal{X}: \delta(\bm{x})=0\}$. Now we define a distance function $\bm{x}\mapsto \tau_{\bm{x}}$ by
\begin{equation*}
\tau_{\bm{x}} = \begin{cases}
\tilde{d}(\bm{x},\mathcal{X}^0\cup\mathcal{X}^+) & \text{if } \quad \bm{x}\in \mathcal{X}^-,\\
\tilde{d}(\bm{x},\mathcal{X}^0\cup\mathcal{X}^-) & \text{if } \quad \bm{x}\in \mathcal{X}^+,\\
0 & \text{if } \quad \bm{x}\in \mathcal{X}^0,
\end{cases}
\end{equation*}
where $\tilde{d}(\bm{x},A)$ denotes the distance of $\bm{x}$ to a set $A$ with respect to the Euclidean norm. Now we present the geometric noise condition for distributions.
\begin{definition} \label{def:geometricnoise}
Let $\mathcal{X}\subset\mathbb{R}^p$ be compact and $P$ be a probability measure of $(\bm{X},A,R)$ on $\mathcal{X}\times\mathcal{A}\times\mathcal{M}$. We say that $P$ has geometric noise exponent $q>0$ if there exists a constant $C>0$ such that
\begin{equation} \label{eq:geometric}
\int_{\mathcal{X}}exp\left(-\frac{\tau^2_{\bm{x}}}{t}\right)|\delta(\bm{x})|\mu(d\bm{x})\leq Ct^{qp/2}, \quad t>0,
\end{equation}
where $\mu$ is the marginal measure on $\bm{X}$.
\end{definition}

The integral condition (\ref{eq:geometric}) describes the concentration of the measure $|\delta(\bm{x})|d\mu$ near the decision boundary in the sense that the less the measure is concentrated in this region the larger the geometric noise exponent can be chosen. The following lemma shows that the geometric noise condition can be used to guarantee the approximation error assumption (\ref{approximateassumption}) when the parameter $\sigma$ of the Gaussian RBF kernel $K_{\sigma}$ is appropriately chosen.

\begin{lemma} \label{thm:geometricbound}
Let $\mathcal{X}$ be the closed unit ball of the Euclidean space $\mathbb{R}^p$, and $P$ be a distribution on $\mathcal{X}\times\mathcal{A}\times\mathcal{M}$ that has geometric noise exponent $0<q<\infty$ with constant $C$ in (\ref{eq:geometric}). Let $\sigma=\lambda^{-\frac{1}{(q+1)p}}$. Then for the Gaussian RBF kernel $K_{\sigma}$,  there is a constant $c>0$ depending only on the dimension $p$, the geometric noise exponent $q$ and constant $C$, such that for all $\lambda>0$ we have
\begin{equation*}
\mathcal{A}(\lambda) \leq c\lambda^{q/(q+1)}.
\end{equation*}
\end{lemma}

The proof provided in Appendix follows the idea from Theorem 2.7 in \citet{Steinwart:FastRateGaussian} by replacing their $2\eta(\bm{x})-1$ with $\delta(\bm{x})$. The key point is the following inequality, for all measurable $f$,
\begin{equation*}
\mathcal{R}_{T,g}(f)-\mathcal{R}^*_{T,g} \leq 2\mathbb{E}(|\delta(\bm{x})|\cdot|f-d^*|),
\end{equation*}
which is the counterpart of Equation~(24) in \citet{Steinwart:FastRateGaussian}.

Now we are ready to present the convergence rate of RWL with Gaussian RBF kernel by combining  Theorem \ref{thm:convergencerate} and Lemma \ref{thm:geometricbound}:
\begin{proposition} \label{thm:gaussianrate}
Let $\mathcal{X}$ be the closed unit ball of the Euclidean space $\mathbb{R}^p$, and $P$ be a distribution on $\mathcal{X}\times\mathcal{A}\times\mathcal{M}$ that has geometric noise exponent $0<q<\infty$ with constant $C$ in (\ref{eq:geometric}). For a measurable function $g$, assume that $\frac{|R-g(\bm{X})|}{\pi(A,\bm{X})}\leq M_0$ almost surely. Consider a Gaussian RBF kernel $K_{\sigma_n}$. Take $\lambda_n=n^{-\frac{q}{3q+1}}$ and $\sigma_n=\lambda_n^{-\frac{1}{(q+1)p}}$. Then with probability at least $1-\delta$,
\begin{equation*}
\mathcal{R}(sign(f_{D_n,\lambda_n})) - \mathcal{R}^* \leq \tilde{c}\sqrt{\log(4/\delta)}n^{-\frac{q}{3q+1}},
\end{equation*}
where the constant $\tilde{c}$ is independent of $\delta$ and $n$, and $\tilde{c}$ decreases as $M_0$ decreases.
\end{proposition}
The optimal rate for the risk is about $n^{-1/3}$ when the geometric noise exponent $q$ is sufficiently large. A better rate may be achieved by using techniques in \citet{Steinwart:FastRateGaussian}, but it is beyond the scope of this work.
\section{Variable selection for RWL} \label{sec:rwlvs}

Variable selection is an important area in modern statistical research.  \citet{Gunter2011:Qualitative} distinguished prescriptive covariates from predictive covariates. The latter refers to covariates which are related to the prediction of outcomes, and the former refers to covariates which help to prescribe optimal ITRs. Clearly, prescriptive covariates are predictive too. In practice, a large number of clinical covariates are often available for estimating optimal ITRs. However, many of them might not be prescriptive. Thus careful variable selection could lead to better performance.

\subsection{Variable selection for linear RWL}
There is a vast body of literature on variable selection for linear classification in statistical learning. For instance, \citet{Zhu2003:L1SVM} and \citet{Fung2004:L1SVM}
investigated the support vector machine (SVM) using the $\ell_1$-norm penalty \citep{Tibshirani1994:lasso}. \citet{Wang2008:hhsvm} applied the elastic-net penalty \citep{Zou2005:elasticnet} for variable selection with an SVM-like method. In this article, we replace the $\ell_2$-norm penalty in RWL with the elastic-net penalty,
\begin{equation*}
\lambda_1||\bm{w}||_1 + \frac{\lambda_2}{2}\bm{w}^T\bm{w},
\end{equation*}
where $||\bm{w}||_1=\sum_{j=1}^p|w_j|$ is the $\ell_1$-norm. As a hybrid of the $\ell_1$-norm and $\ell_2$-norm penalties, the elastic-net penalty retains the variable selection features of
the $\ell_1$-norm penalty, and tends to provide similar estimated coefficients for highly correlated variables, \textit{i.e.} the grouping effect, as the $\ell_2$-norm penalty does. Hence, highly correlated variables are selected or removed together.

The elastic-net penalized linear RWL aims to minimize
\begin{equation*}
\lambda_1||\bm{w}||_1 + \frac{\lambda_2}{2}\bm{w}^T\bm{w} + \frac{1}{n}\sum_{i=1}^{n}\frac{\hat{r}_i}{\pi(a_i,\bm{x}_i)}T\big(a_i(\bm{w}^T\bm{x}_i+b)\big),
\end{equation*}
where $\lambda_1(>0)$ and $\lambda_2(\geq0)$ are regularization parameters.
We still apply the d.c. algorithm to solve the above optimization problem. Following a similar decomposition, the convex subproblem at the $(t + 1)$'th
iteration of the d.c. algorithm is as follows,
\begin{equation} \label{eq:prwllinearsub}
\min_{\bm{w},b} \quad \lambda_1||\bm{w}||_1+\frac{\lambda_2}{2}\bm{w}^T\bm{w} + \frac{1}{n}\sum_{i=1}^n\Big[\phi_1(u_i)\mathbb{I}(\hat{r}_i>0)+\phi_{0}(u_i)\mathbb{I}(\hat{r}_i<0)\Big]\frac{|\hat{r}_i|}{\pi(a_i,\bm{x}_i)}+\sum_{i=1}^{n}\beta_i^{(t)}u_i,
\end{equation}
where $u_i = a_i(\bm{w}^T\bm{x}_i+b)$, and $\beta_i^{(t)}$ is computed by (\ref{eq:beta}). The objective function is a sum of a smooth function and a non-smooth $\ell_1$-norm penalty. Since L-BFGS can only deal with smooth objective functions, it may not work here. Instead we use projected scaled sub-gradient (PSS) algorithms \citep{Schmidt2010:LASSO}, which are extensions of L-BFGS to the case of optimizing a smooth function with an $\ell_1$-norm penalty. Several PSS algorithms are proposed in \citet{Schmidt2010:LASSO}. Among them, the Gafni-Bertseka variant is particularly effective for our purpose. After solving the subproblem, we update $\beta_i^{(t+1)}$ by (\ref{eq:beta}). The procedure is repeated until $\bm{\beta}$ converges. The obtained decision function is $\hat{f}(\bm{x})=\hat{\bm{w}}^T\bm{x}+\hat{b}$, and thus the estimated optimal ITR is the sign of $\hat{f}(\bm{x})$.

\subsection{Variable selection for RWL with nonlinear kernels}
Many researchers have noticed that nonlinear kernel methods may perform poorly when there are irrelevant covariates presented in the data \citep{Weston2000:SVMSelection, Grandvalet2002:FeatureSelection, Lin2006:COSSO, Lafferty2008:Rodeo}. For optimal treatment regimes, we may suffer a similar problem when using a nonlinear kernel. For the Gaussian RBF kernel, we assume that the geometric noise condition in Definition \ref{def:geometricnoise} is satisfied with exponent $q$. When several additional non-descriptive covariates are included in the data, the geometric noise exponent is decreased according to (\ref{eq:geometric}), and hence so is the convergence rate in Proposition \ref{thm:gaussianrate}. The presence of non-descriptive covariates thus can deteriorate the performance of RWL.

The variable selection approach proposed in this section is inspired by the KNIFE algorithm \citep{Allen2013:KNIFE}, where a set of scaling factors on the covariates are employed to form a regularized loss function. The idea of scaling covariates within the kernel has appeared in several methods from the machine learning community \citep{Weston2000:SVMSelection, Grandvalet2002:FeatureSelection, Rakotomamonjy2003:svm, Argyriou2006:kernelselection}.

We take the Gaussian RBF kernel as an example. Define the covariates-scaled Gaussian RBF kernel,
\begin{equation*}
K_{\bm{\eta}}(\bm{x},\bm{z}) = \exp\left(-\sum_{j=1}^{p}\eta_j(x_j-z_j)^2\right),
\end{equation*}
where $\bm{\eta}=(\eta_1,\cdots,\eta_p)^T\geq\bm{0}$. Here each covariate $x_j$ is scaled by $\sqrt{\eta_j}$. Setting ${\eta_j}=0$ is equivalent to discarding the $j$'th covariate. The hyperparameter $\sigma$ in the original Gaussian RBF kernel is absorbed to the scaling factors. Similar with KNIFE, we seek $(\hat{\bm{v}}, \hat{b}, \hat{\bm{\eta}})$ to minimize the following optimization problem:
\begin{eqnarray} \label{nonlinearpenalizedrwlprime}
\min_{\bm{v},b,\bm{\eta}} &&\lambda_1||\bm{\eta}||_1+ \frac{\lambda_2}{2}\sum_{i,j=1}^nv_iv_jK_{\bm{\eta}}(\bm{x}_i,\bm{x}_j) \nonumber \\
&&\qquad + \frac{1}{n}\sum_{i=1}^{n}\frac{\hat{r}_i}{\pi(a_i,\bm{x}_i)}T\Big(a_i\big(\sum_{j=1}^nv_jK_{\bm{\eta}}(\bm{x}_i,\bm{x}_j)+b\big)\Big), \\
\textrm{subject to} && \bm{\eta} \geq \bm{0}, \nonumber
\end{eqnarray}
where $\lambda_1(>0)$ and $\lambda_2(>0)$ are regularization parameters.
Compared with previous nonlinear RWL optimization problem in (\ref{nonlinearrwlprime}), (\ref{nonlinearpenalizedrwlprime}) has an $\ell_1$-norm penalty on scaling factors. The objective function is singular at $\bm{\eta}=\bm{0}$ due to nonnegativity of scaling factors. So (\ref{nonlinearpenalizedrwlprime}) may produce zero solutions for some of the $\bm{\eta}$, and hence performs variable selection.

We adopt a similar decomposition trick that has been used several times before. The subproblem at the $(t+1)$'th iteration is
\begin{eqnarray}
\min_{\bm{v},b,\bm{\eta}} &&\lambda_1||\bm{\eta}||_1+\frac{\lambda_2}{2}\sum_{i,j=1}^nv_iv_jK_{\bm{\eta}}(\bm{x}_i,\bm{x}_j) \nonumber \\
&&\qquad + \frac{1}{n}\sum_{i=1}^n\Big[\phi_1(u_i)\mathbb{I}(\hat{r}_i>0)+\phi_{0}(u_i)\mathbb{I}(\hat{r}_i<0)\Big]\frac{|\hat{r}_i|}{\pi(a_i,\bm{x}_i)}+\sum_{i=1}^{n}\beta_i^{(t)}u_i , \label{eq:prwlnonlinearsub}\\
\textrm{subject to} && \bm{\eta}\geq 0, \nonumber
\end{eqnarray}
where $u_i = a_i\big(\sum_{j=1}^nv_jK_{\bm{\eta}}(\bm{x}_i,\bm{x}_j)+b\big)$, and $\beta_i^{(t)}$ is computed by (\ref{eq:beta}). Recall that the d.c. algorithm is a special case of the Majorize-Minimization (MM) algorithm. Although with the covariate-scaled kernel the subproblem is nonconvex, we still apply a similar iterative procedure as in the d.c. algorithm, \textit{i.e.} solving a sequence of subproblems (\ref{eq:prwlnonlinearsub}), but based on the MM algorithm. The subproblem (\ref{eq:prwlnonlinearsub}) is a smooth optimization problem with box constraints. In this article, we use L-BFGS-B \citep{Byrd1995:LBFGSB, Morales2011:LBFGSB}, which extends L-BFGS to handle simple box constraints on variables. After solving the subproblem (\ref{eq:prwlnonlinearsub}), we update $u_i$ and $\bm{\beta}$. The procedure is repeated until convergence. Then the obtained decision function is $\hat{f}(\bm{x})=\sum_{i=1}^n\hat{v}_iK_{\hat{\bm{\eta}}}(\bm{x}_i,\bm{x})+\hat{b}$, and thus the estimated optimal ITR is the sign of $\hat{f}(\bm{x})$. The covariates with nonzero scaling factors $\hat{\eta}$ are identified to be important in estimating ITRs.

This variable selection technique can be applied to other nonlinear kernels, such as the polynomial kernel, and even to the linear kernel. However for the linear kernel, we recommend the approach in the prior section, where the subproblem (\ref{eq:prwllinearsub}) is a convex optimization problem with $p+1$ variables. Note that the subproblem (\ref{eq:prwlnonlinearsub}) in this section is nonconvex even for the linear kernel, and there are $n+p+1$ variables for the optimization problem. This is much more challenging.

\section{Simulation studies}
We carried out extensive simulation studies to investigate finite sample performance of the proposed RWL methods.

\begin{figure}[tbp]
\centering
\subfigure[Scenario 1]{\label{fig:linear} \includegraphics[scale=0.5]{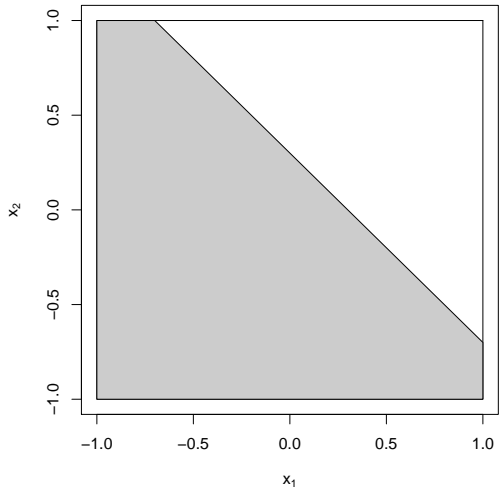}} 
\subfigure[Scenario 2]{\label{fig:parabola} \includegraphics[scale=0.5]{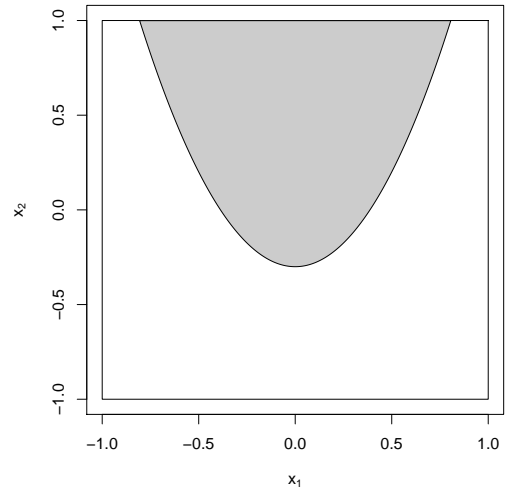}} \\
\subfigure[Scenario 3]{\label{fig:circle} \includegraphics[scale=0.5]{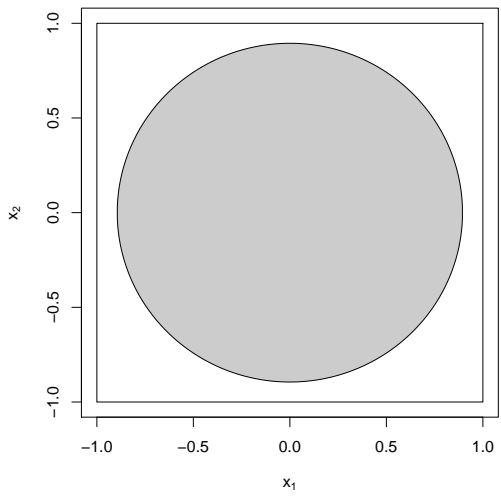}} 
\subfigure[Scenario 4]{\label{fig:ring} \includegraphics[scale=0.5]{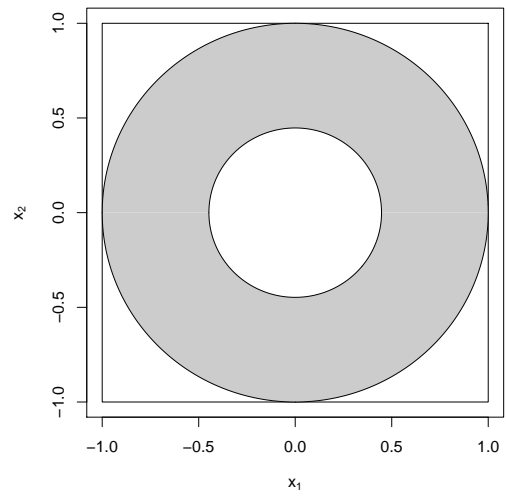}}
\caption{True optimal ITRs in simulation studies. For subjects in the shade area, the best treatment is 1; for subjects in the white area, the best treatment is $-1$.} \label{fig:decisionboundary}
\end{figure}

We first evaluated the performance of RWL methods with low-dimensional covariates. In the simulations, we generated 5-dimensional vectors of clinical covariates $x_1,\cdots,x_5$, consisting of independent uniform random variables $U(-1,1)$. The treatment $A$ was generated from $\{-1,1\}$ independently of $X$ with $P(A=1)=0.5$. That is, $\pi(a,\bm{x})=0.5$ for all $a$ and $\bm{x}$. The response $R$ was normally distributed with mean $Q_0=\mu_0(\bm{x})+\delta_0(\bm{x})\cdot a$ and standard deviation 1, where $\mu_0(\bm{x})$ is the common effect for clinical covariates, and $\delta_0(\bm{x})\cdot a$ is the interaction between treatment and clinical covariates. We considered five scenarios with different choices of $\mu_0(\bm{x})$ and $\delta_0(\bm{x})$:
\begin{enumerate}
\item[(0)] $\mu_0(\bm{x})=1+x_1+x_2+2x_3+0.5x_4$; { } $\delta_0(\bm{x})=0.4(x_2-0.25x_1^2-1)$.
\item[(1)] $\mu_0(\bm{x})=1+x_1+x_2+2x_3+0.5x_4$; { } $\delta_0(\bm{x})=1.8(0.3-x_1-x_2)$.
\item[(2)] $\mu_0(\bm{x})=1+x_1+x_2+2x_3+0.5x_4$; { } $\delta_0(\bm{x})=1.3(x_2 - 2x_1^2 +0.3)$.
\item[(3)] $\mu_0(\bm{x})=1+x_1^2+x_2^2+2x_3^2+0.5x_4^2$; { } $\delta_0(\bm{x})=3.8(0.8 - x_1^2 - x_2^2)$.
\item[(4)] $\mu_0(\bm{x})=1+x_1^2+x_2^2+2x_3^2$; { } $\delta_0(\bm{x})=10(1 - x_1^2 - x_2^2)(x_1^2+x_2^2-0.2)$.
\end{enumerate}
Scenario 0 is similar with the second scenario in \citet{Zhao:OWL2012}. When $x_2$ is restricted to $[-1,1]$, the treatment arm $-1$ is always better than the treatment arm 1 on average. Thus the true optimal ITR in Scenario 0 is to assign all subjects to arm $-1$. The decision boundaries for the remaining four scenarios are determined by $x_1$ and $x_2$ only. The scenarios have different decision boundaries in truth. The decision boundary is a line in Scenario 1, a parabola in Scenario 2, a circle in Scenario 3, and a ring in Scenario 4. Their true optimal ITRs are illustrated in Figure~\ref{fig:decisionboundary}.
It is unclear how often a circle or ring boundary structure will occur in practice, although general nonlinear boundaries are likely to arise frequently. The purpose of including these in the simulations is to verify that the proposed approach can handle even the most difficult boundary structures.
The coefficients in Scenarios $0\sim3$ were chosen to reflect a medium effect size according to Cohen's d index \citep{Cohen:Power}; and the coefficients in Scenario 4 reflect a small effect size. The Cohen's d index is defined as the standardized difference in mean responses between two treatment arms, that is,
\begin{equation*}
es = \frac{|\mathbb{E}(R|A=1)-\mathbb{E}(R|A=-1)|}{\sqrt{[Var(R|A=1)+Var(R|A=-1)]/2}}.
\end{equation*}

We compared the performance of the following five methods:
\begin{itemize}
\item[(1)] The proposed RWL using the Gaussian RBF kernel (RWL-Gaussian). Residuals were estimated by a linear main effects model.
\item[(2)] The proposed RWL using the linear kernel (RWL-Linear). Residuals were estimated by a linear main effects model.
\item[(3)] OWL proposed in \citet{Zhao:OWL2012} using the Gaussian RBF kernel (OWL-Gaussian).
\item[(4)] OWL using the linear kernel (OWL-Linear).
\item[(5)] $\ell_1$-PLS proposed by \citet{Qian:ITR2011}.
\end{itemize}
The OWL methods were reviewed in the method section. Since the OWL methods can only handle nonnegative outcomes, we subtracted $\min\{r_i\}$ from all outcome responses. This is essentially the approach used in \citet{Zhao:OWL2012} [personal communication].
In the simulation studies, $\ell_1$-PLS approximated $\mathbb{E}(R|\bm{X},A)$ using the basis function set $(1,\bm{X},A,\bm{X}A)$, and applied LASSO \citep{Tibshirani1994:lasso} for variable selection. The optimal ITR was determined by the sign of the difference between the estimated $\mathbb{E}(R|\bm{X},A=1)$ and $\mathbb{E}(R|\bm{X},A=-1)$.

For each scenario, we considered two sample sizes for training datasets: $n=100$ and $n=400$, and repeated the simulation 500 times. All five methods had at least one tuning parameter. For example, there was one tuning parameter, $\lambda$, in RWL-linear; and two tuning parameters, $\lambda$ and $\sigma$ in RWL-Gaussian. In the simulations, we applied a 10-fold cross-validation procedure to tune parameters. For RWL-Linear, we searched over a pre-specified finite set of $\lambda$'s to select the best one maximizing the average of estimated values from validation data; while for RWL-Gaussian, we searched over a pre-specified finite set of $(\lambda, \sigma)$'s to select the best pair. The selection of tuning parameters in $\ell_1$-PLS and OWL followed similarly.
In the comparisons, the performances of five methods were evaluated by two criteria: the first criterion is the value function under the estimated optimal ITR when we applied to an independent and large test dataset; the second criterion is the misclassification error rate of the estimated optimal ITR from the true optimal ITR on the large test dataset.
Specifically, a test set with 10,000 observations was simulated to assess the performance. The estimated value function under any ITR $d$ is given by $\mathbb{P}_n^*[R\mathbb{I}(A=d(\bm{X}))/\pi(A,\bm{X})]/\mathbb{P}_n^*[\mathbb{I}(A=d(\bm{X}))/\pi(A,\bm{X})]$ \citep{Murphy:Design2005}, where $\mathbb{P}_n^*$ denotes the empirical average using the test data; the misclassification rate under any ITR $d$ is given by
$\mathbb{P}_n^*[\mathbb{I}(d(\bm{X})\neq d^*(\bm{X}))]$, where $d^*$ is the true optimal ITR which is known when generating the simulated data.

The simulation results are presented in Table~\ref{tab:valuecontinuous}. We report the mean and standard deviation of value functions and misclassification rates over 500 replications. In Scenario 0, the true optimal ITR would assign all subjects to treatment arm $-1$. When the sample size was small $(n=100)$, $\ell_1$-PLS, OWL-Linear and RWL-Linear showed similar performance, while OWL-Gaussian and RWL-Gaussian were slightly worse. Note that when $x_2$ can go beyond 1, the heterogeneous treatment effect does exist with non-linear (parabola) decision boundary. The deteriorated performance of OWL-Gaussian and RWL-Gaussian may be due to unexpected extrapolation. When the sample size was large $(n=400)$, all methods performed similarly, and assigned almost all subjects to arm $-1$.
Scenario 1 was a linear example. The model specification in $\ell_1$-PLS was correct, and this method performed very well. The proposed RWL methods, with either the linear kernel or the Gaussian RBF kernel, yielded similar performance with $\ell_1$-PLS, and were much better than OWL methods.
Another advantage of using RWL versus OWL was also reflected by a smaller variance in value functions. The idea of using residuals instead of the original outcomes is able to stabilize the variance, as we discussed in the method section.

The conditional mean outcomes and decision boundaries in the remaining three scenarios were nonlinear. $\ell_1$-PLS, RWL-Linear, and OWL-Linear were misspecified, and hence they did not perform very well in these three scenarios. We focus on the comparison between RWL-Gaussian and OWL-Gaussian. These two methods equipped with the Gaussian RBF kernel should have the ability to capture the nonlinear structure of decision boundaries. In Scenarios 2 and 3, RWL-Gaussian outperformed OWL-Gaussian in terms of achieving higher values and smaller variances. It was challenging to find the optimal ITR in Scenario 4 since the decision boundary was complicated and the effective size was small. When the sample size was small ($n=100$), the performance of RWL-Gaussian was only slightly better than that of OWL-Gaussian, but with larger variance. It might be too hard to learn the complicated decision boundary from the limited sample.
However, when the sample size increased, RWL-Gaussian showed significantly better performance than other methods. The excellent performance of RWL-Gaussian confirms its power in finding the optimal ITR. Although using the Gaussian RBF kernel, the performance of OWL-Gaussian was not comparable with that of RWL-Gaussian. In Scenario 2, OWL-Gaussian performed even worse than $\ell_1$-PLS and RWL-Linear, both of which can only detect linear boundaries. While in Scenario 4, OWL-Gaussian gave similar performance as OWL-Linear. \citet{Zhao:OWL2012} also reported similar finding in their simulation studies that there are no large differences in the performance between OWL-Linear and OWL-Gaussian.

As we demonstrated in Section~\ref{sec:rwl}, OWL methods tend to favor the treatment assignments that subjects actually received, and this behavior deteriorates finite sample performance. We calculated treatment matching factors of these five methods on the training data. They are shown in Table~\ref{tab:factorcontinuous}. We expect the treatment matching factor to be close to 1. However, the treatment matching factors for OWL methods were greater than 1. OWL-Gaussian achieved the largest treatment matching factors, especially when the sample size was small. This may partially explain why OWL-Gaussian may not outperform OWL-Linear even when the decision boundary is nonlinear. Whereas for RWL methods, the treatment matching factors were close to 1. By appropriately controlling the treatment matching factors, the aim of RWL methods is to maximize the estimate in (\ref{eq:empiricalreward2}), and hence to improve finite sample performance.

We applied a linear main effects model to estimate residuals. The model was correctly specified for Scenarios 0, 1 and 2, and misspecified for Scenarios 3 and 4. However, the superior performance of RWL methods in Scenarios 3 and 4 demonstrates the robustness of our methods to residual estimates. We also considered a null model, which was misspecified for all scenarios, to estimate residuals. The results were only slightly worse than those shown in Table~\ref{tab:valuecontinuous} for Scenarios 1 and 2 when $n=100$, and other scenarios were similar especially when the sample size increased [results not shown].

\begin{table}[tbp]
\centering
\caption{Mean (std) of empirical value functions and misclassification rates evaluated on independent test data for 5 simulation scenarios with 5 covariates. The best value function and minimal misclassification rate for each scenario and sample size combination are in bold.}
\label{tab:valuecontinuous}
\begin{tabular}{@{}lccccc@{}}
\addlinespace
\toprule
& \multicolumn{2}{c}{$n=100$} & \phantom{a} & \multicolumn{2}{c}{$n=400$} \\
\cmidrule{2-3} \cmidrule{5-6}
& Value & Misclassification & & Value & Misclassification \\
\midrule
\multicolumn{6}{c}{Scenario 0 (Optimal value $1.43$)} \\
\midrule
$\ell_1$-PLS & \textbf{1.41 (0.07)} & \textbf{0.05 (0.10)} & & 1.42 (0.01) & 0.02 (0.04) \\
OWL-Linear & 1.39 (0.11) & 0.07 (0.14) & & \textbf{1.43 (0.02)} & \textbf{0.01 (0.04)} \\
OWL-Gaussian & 1.33 (0.15) & 0.13 (0.18) & & 1.41 (0.05) & 0.03 (0.08) \\
RWL-Linear & 1.40 (0.05) & 0.06 (0.09) & & 1.42 (0.02) & 0.03 (0.05) \\
RWL-Gaussian & 1.34 (0.10) & 0.13 (0.13) & & 1.40 (0.04) & 0.07 (0.07)\\
\midrule
\multicolumn{6}{c}{Scenario 1 (Optimal value $2.25$)} \\
\midrule
$\ell_1$-PLS & \textbf{2.22 (0.03)} & \textbf{0.06 (0.04)} & & \textbf{2.24 (0.02)} & \textbf{0.03 (0.03)}\\
OWL-Linear & 1.91 (0.22) & 0.23 (0.09) & & 2.08 (0.14) & 0.16 (0.06) \\
OWL-Gaussian & 1.88 (0.24) & 0.24 (0.09) & & 2.08 (0.12) & 0.16 (0.06)\\
RWL-Linear & 2.19 (0.04) & 0.09 (0.04) & & 2.23 (0.01) & 0.05 (0.02) \\
RWL-Gaussian & 2.17 (0.06) & 0.11 (0.04) & & 2.22 (0.02) & 0.05 (0.02) \\
\midrule
\multicolumn{6}{c}{Scenario 2 (Optimal value $1.96$)} \\
\midrule
$\ell_1$-PLS & 1.71 (0.07) & 0.24 (0.04) & & 1.75 (0.01) & 0.22 (0.01)\\
OWL-Linear & 1.51 (0.12) & 0.32 (0.05) & & 1.59 (0.10) & 0.29 (0.06)\\
OWL-Gaussian & 1.49 (0.15) & 0.33 (0.06) & & 1.63 (0.11) & 0.26 (0.05)\\
RWL-Linear & 1.66 (0.08) & 0.26 (0.04) & & 1.74 (0.03) & 0.23 (0.02) \\
RWL-Gaussian & \textbf{1.75 (0.09)} & \textbf{0.20 (0.05)} & & \textbf{1.90 (0.03)} & \textbf{0.10 (0.03)}\\
\midrule
\multicolumn{6}{c}{Scenario 3 (Optimal value $3.88$)} \\
\midrule
$\ell_1$-PLS & 3.00 (0.11) & 0.38 (0.03) & & 3.03 (0.04) & 0.37 (0.01)\\
OWL-Linear & 2.98 (0.10) & 0.38 (0.03) & & 3.01 (0.02) &  0.38 (0.01) \\
OWL-Gaussian & 3.21 (0.18) & 0.32 (0.05) & & 3.56 (0.12) & 0.22 (0.04) \\
RWL-Linear & 3.15 (0.13) & 0.34 (0.03) & & 3.28 (0.04) & 0.31 (0.01) \\
RWL-Gaussian & \textbf{3.62 (0.12)} & \textbf{0.19 (0.04)} & & \textbf{3.82 (0.04)} & \textbf{0.10 (0.02)} \\
\midrule
\multicolumn{6}{c}{Scenario 4 (Optimal value $3.87$)} \\
\midrule
$\ell_1$-PLS & 2.29 (0.14) & 0.49 (0.08) & & 2.38 (0.15) & 0.55 (0.08)\\
OWL-Linear & 2.42 (0.14) & 0.55 (0.08) & & 2.49 (0.11) & 0.59 (0.06)\\
OWL-Gaussian &  2.43 (0.15) & 0.52 (0.07) & & 2.59 (0.15) & 0.50 (0.07) \\
RWL-Linear & 2.42 (0.14) & 0.54 (0.09) & & 2.49 (0.10) & 0.58 (0.08) \\
RWL-Gaussian &  \textbf{2.68 (0.28)} & \textbf{0.43 (0.08)} & & \textbf{3.49 (0.08)} & \textbf{0.22 (0.03)} \\
\bottomrule
\end{tabular}
\vfill
\end{table}

\begin{table}[tbp]
\centering
\caption{Mean (std) of treatment matching factors evaluated on the training data for 5 simulation scenarios with 5 covariates. }
\label{tab:factorcontinuous}
\begin{tabular}{lcc}
\addlinespace
\toprule
& $n=100$ & $n=400$ \\
\midrule
\multicolumn{3}{c}{Scenario 0} \\
\midrule
$\ell_1$-PLS & 1.00 (0.04) & 1.00 (0.01) \\
OWL-Linear & 1.03 (0.07) &  1.00 (0.01) \\
OWL-Gaussian & 1.14 (0.24) & 1.02 (0.06) \\
RWL-Linear & 1.00 (0.04) & 1.00 (0.01)  \\
RWL-Gaussian & 1.00 (0.06) & 1.00 (0.03)  \\
\midrule
\multicolumn{3}{c}{Scenario 1} \\
\midrule
$\ell_1$-PLS &  0.99 (0.10) &  0.99 (0.05)   \\
OWL-Linear &  1.10 (0.08) &  1.04 (0.04)   \\
OWL-Gaussian &  1.15 (0.13) &  1.05 (0.05)  \\
RWL-Linear &  0.99 (0.09) &  0.99 (0.04)   \\
RWL-Gaussian &  0.99 (0.08) &  0.99 (0.04)   \\
\midrule
\multicolumn{3}{c}{Scenario 2} \\
\midrule
$\ell_1$-PLS &  1.00 (0.09) &  1.00 (0.04)  \\
OWL-Linear &  1.06 (0.09) &  1.03 (0.04)   \\
OWL-Gaussian &  1.18 (0.20) &  1.10 (0.07)  \\
RWL-Linear &  1.00 (0.08) &  1.00 (0.04)  \\
RWL-Gaussian &  1.01 (0.07) & 1.00 (0.04)  \\
\midrule
\multicolumn{3}{c}{Scenario 3} \\
\midrule
$\ell_1$-PLS &  1.00 (0.05) &  1.00 (0.01)    \\
OWL-Linear &  1.02 (0.06) &  1.00 (0.02)    \\
OWL-Gaussian & 1.32 (0.17) &  1.14 (0.05)   \\
RWL-Linear &  1.01 (0.06) &  1.01 (0.04)   \\
RWL-Gaussian &  1.04 (0.06) &  1.02 (0.04)   \\
\midrule
\multicolumn{3}{c}{Scenario 4} \\
\midrule
$\ell_1$-PLS &  1.00 (0.08) &  1.00 (0.03)   \\
OWL-Linear &  1.07 (0.10) &  1.02 (0.04)  \\
OWL-Gaussian &  1.37 (0.34) & 1.27 (0.24)    \\
RWL-Linear &  1.00 (0.06) &  1.00 (0.03)  \\
RWL-Gaussian &  1.00 (0.08) & 1.02 (0.04) \\
\bottomrule
\end{tabular}
\end{table}

We then evaluated the performance of RWL methods with moderate-dimensional clinical covariates. We adopted the same data generating procedure as in the above five scenarios except that the dimension of clinical covariates was increased from 5 to 50 by adding 45 independent uniform random variables $U(-1,1)$. Thus among all these 50 clinical covariates, only $x_1$ and $x_2$ are attributed to the decision boundaries for Scenarios $1\sim4$. We repeated the simulations, and the results are shown in Table~\ref{tab:valuecontinuous50}. In Scenario 0, all methods performed similar, and assigned almost all subjects to the treatment arm $-1$ when the sample size was large $(n=400)$. In Scenario 1, $\ell_1$-PLS performed very well because of correct model specification and inside variable selection techniques. Although RWL methods showed better performance than OWL methods, they were not comparable to $\ell_1$-PLS. The decision boundaries in Scenarios~$2\sim4$ are nonlinear. Both RWL and OWL failed to detect the decision boundaries in these scenarios. They produced similar performance with either the linear kernel or Gaussian RBF kernel. RWL and OWL methods with Gaussian RBF kernel may perform particularly poorly when many non-descriptive covariates are present.

We tested the performance of variable selection approaches described in Section 4 for the linear kernel and Gaussian RBF kernel. They are called RWL-VS-Linear and RWL-VS-Gaussian, respectively, in this article.
There were two tuning parameters, $\lambda_1$ and $\lambda_2$, for both approaches. We applied a 10-fold cross-validation procedure to tune parameters from a pre-specified finite set of $(\lambda_1, \lambda_2)$'s. The simulation results are also shown in Table~\ref{tab:valuecontinuous50}. In Scenario 0, both RWL-VS methods showed similar performance as RWL methods. In Scenario 1, both RWL-VS-Linear and RWL-VS-Gaussian improved the performance of RWL methods. When the sample size was 400, the performance was as good as that of $\ell_1$-PLS. For Scenarios~$2\sim4$, we only present results from RWL-VS-Gaussian. When the sample size was small ($n=100$), the performance was slightly better (in Scenarios 2 and 3) than, or similar (in scenario 4) as that of RWL-Gaussian. It is well known that large sample sizes are needed to find interactions in models. So for similar reasons, we may expect large samples to be necessary to find and confirm the existence of heterogeneity of treatment effects, and accurately define the decision boundary. In the simulations, when the sample size increased, the performance of RWL-VS-Gaussian improved significantly. Variable selection is necessary for RWL when there are many non-descriptive clinical covariates.

\begin{table}[tbp]
\centering
\small
\caption{Mean (std) of empirical value functions and misclassification rates evaluated on independent test data for 5 simulation scenarios with 50 covariates. The best value function and minimal misclassification rate for each scenario and sample size combination are in bold.}
\label{tab:valuecontinuous50}
\begin{tabular}{@{}lccccc@{}}
\addlinespace
\toprule
& \multicolumn{2}{c}{$n=100$} & \phantom{a} & \multicolumn{2}{c}{$n=400$} \\
\cmidrule{2-3} \cmidrule{5-6}
& Value & Misclassification & & Value & Misclassification \\
\midrule
\multicolumn{6}{c}{Scenario 0 (Optimal value $1.43$)} \\
\midrule
$\ell_1$-PLS & 1.34 (0.15) & 0.11 (0.17) & & 1.42 (0.03) & 0.03 (0.06)\\
OWL-Linear & 1.36 (0.15) & 0.08 (0.17) & & \textbf{1.43 (0.03)} & \textbf{0.01 (0.03)} \\
OWL-Gaussian & 1.35 (0.14) & 0.10 (0.16) & & 1.42 (0.04) & \textbf{0.01 (0.05)} \\
RWL-Linear & \textbf{1.38 (0.11)} & \textbf{0.07 (0.13)} & & 1.42 (0.04) & 0.03 (0.06)\\
RWL-Gaussian & 1.34 (0.13) & 0.11 (0.15) & & 1.40 (0.05) & 0.04 (0.07) \\
\hdashline
RWL-VS-Linear & 1.34 (0.11) & 0.11 (0.14) & & 1.40 (0.04) & 0.05 (0.06) \\
RWL-VS-Gaussian & 1.33 (0.12) & 0.13 (0.14) & & 1.41 (0.03) & 0.05 (0.06) \\
\midrule
\multicolumn{6}{c}{Scenario 1 (Optimal value $2.25$)} \\
\midrule
$\ell_1$-PLS & \textbf{2.17 (0.11)} & \textbf{0.09 (0.06)} && \textbf{2.23 (0.02)} & \textbf{0.04 (0.03)} \\
OWL-Linear & 1.51 (0.12) & 0.37 (0.03) && 1.68 (0.10) & 0.32 (0.03)\\
OWL-Gaussian & 1.49 (0.13) & 0.38 (0.04) && 1.65 (0.11) & 0.32 (0.04)\\
RWL-Linear & 1.75 (0.14) & 0.29 (0.05) && 2.14 (0.03) & 0.13 (0.02)\\
RWL-Gaussian & 1.77 (0.12) & 0.29 (0.04) && 2.14 (0.03) & 0.13 (0.02)\\
\hdashline
RWL-VS-Linear & 2.05 (0.14) & 0.17 (0.07) && 2.22 (0.02) & 0.06 (0.03)\\
RWL-VS-Gaussian & 2.02 (0.18) & 0.18 (0.08) &&  2.21 (0.08) & 0.06 (0.04)\\
\midrule
\multicolumn{6}{c}{Scenario 2 (Optimal value $1.96$)} \\
\midrule
$\ell_1$-PLS & \textbf{1.56 (0.19)} & \textbf{0.30 (0.07)} && 1.73 (0.03) &  0.23 (0.01) \\
OWL-Linear & 1.42 (0.14) & 0.37 (0.04) && 1.47 (0.06) & 0.35 (0.01) \\
OWL-Gaussian & 1.40 (0.14) & 0.38 (0.04) && 1.47 (0.07) & 0.35 (0.02) \\
RWL-Linear & 1.42 (0.12) & 0.36 (0.04) && 1.60 (0.05) & 0.28 (0.02) \\
RWL-Gaussian & 1.40 (0.14) & 0.36 (0.04) && 1.62 (0.05) & 0.28 (0.02) \\
\hdashline
RWL-VS-Gaussian & \textbf{1.56 (0.15)} & \textbf{0.30 (0.07)} &&  \textbf{1.92 (0.06)} & \textbf{0.09 (0.05)}\\
\midrule
\multicolumn{6}{c}{Scenario 3 (Optimal value $3.88$)} \\
\midrule
$\ell_1$-PLS & 2.90 (0.19) & 0.40 (0.05) && 3.00 (0.04) & 0.38 (0.01)\\
OWL-Linear & 2.96 (0.14) & 0.39 (0.04) && 3.00 (0.02) & 0.38 (0.01) \\
OWL-Gaussian & 2.97 (0.11) & 0.39 (0.03) && 3.00 (0.03) & 0.38 (0.01)\\
RWL-Linear & 2.90 (0.18) & 0.40 (0.05) && 3.00 (0.05) & 0.38 (0.01) \\
RWL-Gaussian & 2.88 (0.18) & 0.41 (0.05)&& 2.96 (0.09) & 0.39 (0.02)\\
\hdashline
RWL-VS-Gaussian & \textbf{3.17 (0.33)} & \textbf{0.33 (0.09)} &&  \textbf{3.82 (0.13)} & \textbf{0.09 (0.05)}\\
\midrule
\multicolumn{6}{c}{Scenario 4 (Optimal value $3.87$)} \\
\midrule
$\ell_1$-PLS & 2.32 (0.09) & \textbf{0.49 (0.05)} && 2.36 (0.09) & 0.52 (0.05) \\
OWL-Linear & 2.41 (0.14) & 0.55 (0.08) && 2.48 (0.11) & 0.59 (0.06) \\
OWL-Gaussian &  \textbf{2.43 (0.13)} & 0.55 (0.07) && 2.47 (0.11) & 0.58 (0.06) \\
RWL-Linear & 2.38 (0.13) & 0.53 (0.07) && 2.44 (0.12) & 0.56 (0.07) \\
RWL-Gaussian &  2.37 (0.13) & 0.53 (0.07) && 2.44 (0.12) & 0.56 (0.07)\\
\hdashline
RWL-VS-Gaussian & 2.37 (0.12) & 0.51 (0.05) &&  \textbf{3.47 (0.47)} & \textbf{0.21 (0.15)}\\
\bottomrule
\end{tabular}
\end{table}

\afterpage{\clearpage}

\section{Data analysis} \label{sec:dataanalysis}

We applied the proposed methods to analyze data from the EPIC randomized clinical trial \citep{Treggiari2011:EPIC}. The trial was designed to determine the optimal anti-pseudomonal treatment strategy in children with cystic fibrosis (CF) who recently acquired Pseudomonas aeruginosa (\textit{Pa}).
Newly identified \textit{Pa} was defined as the first lifetime documented \textit{Pa} positive culture within 6 months of baseline or a \textit{Pa} positive culture within 6 months of baseline after a two-year absence of \textit{Pa} culture-positivity.   Other eligibility criteria have been previously reported \citep{Treggiari2009:EPIC}.
A total of 304 patients ages 1$\sim$12 were randomized in a 1:1 ratio to one of two maintenance treatment strategies: cycled therapy (treatment with anti-pseudomonal therapy provided in quarterly cycles regardless of \textit{Pa} positivity) and culture-based therapy (treatment only in response to identification of \textit{Pa} positive cultures from quarterly cultures).
All patients regardless of randomization strategy received an initial course of anti-pseudomonal therapy in response to their \textit{Pa} positive culture at eligibility and proceeded with treatment according to their randomization strategy.
In this article, we considered two endpoints over the course of the 18-month study. One is the number of \textit{Pa} positive cultures from scheduled follow-up quarterly cultures, and the other is the number of pulmonary exacerbations (PE) requiring any (intravenous, inhaled, or oral) antibiotic use or hospitalization during the study. In the original clinical trial, there was no significant difference between the two maintenance treatment strategies at the population level for either the \textit{Pa} outcome (p-value 0.222) or PE outcome (p-value 0.280).

We considered 10 baseline clinical covariates, including age, gender, F508del genotype, weight, height, BMI, first documented lifetime \textit{Pa} positive culture at eligibility versus positive after a two-year history of negative cultures (neverpapos), use of antibiotics within 6 months prior to baseline (anyabxhist), \textit{Pa} positive at the baseline visit versus \textit{Pa} positive within 6 months prior to baseline only (pabl), and positive versus negative baseline S. aureus culture status (sabl). There were three levels (homozygous, heterozygous and other) for F508del genotype. We coded genotype as two dichotomous covariates, homozygous and heterozygous, each referenced in relation to the ``othe'' genotype category.
Four patients did not give consent to have their data put into the databank. We also excluded 17 patients with missing covariate values from the analyses. The data used here consisted of 283 patients: 141 in the cycled therapy, and 142 in the culture-based therapy. Each clinical covariate was linearly scaled to the range $[-1,+1]$, as recommended in \citet{Hsu2003:svmguide}.

For the \textit{Pa}-related endpoint, we used the ratio of the number of \textit{Pa} negative cultures to the number of \textit{Pa} cultures over the follow up period as the outcome. We were seeking an ITR to reduce the number of \textit{Pa} positive cultures, so larger outcomes were preferred. We compared seven methods: $\ell_1$-PLS, OWL-Linear, OWL-Gaussian, RWL-Linear, RWL-Gaussian, RWL-VS-Linear, and RWL-VS-Gaussian. $\ell_1$-PLS used the basis function set $(1,\bm{X},A,\bm{X}A)$ in the regression model. We treated the outcome as continuous, and used a linear main effects model to estimate residuals for RWL methods. We used a 10-fold cross-validation procedure to tune parameters. The estimated rule was then used to predict the optimal treatment for each patient. The predicted treatment allocation is shown in Table~\ref{tab:epicpaoutcome}.
To evaluate the performance of estimated rules, we again carried out a 10-fold cross validation procedure. The data were randomly partitioned into 10 roughly equal-sized parts. We estimated the ITR on nine parts of the data using the tuned parameters, and predicted optimal treatments on the part left out. We repeated the procedure nine more times to predict the other parts, and obtained the predicted treatment for each patient. The first evaluation criterion was the value function, which is given by $\mathbb{P}_n[R\mathbb{I}(A=Pred)/\pi(A,\bm{X})]/\mathbb{P}_n[\mathbb{I}(A=Pred)/\pi(A,\bm{X})]$, where $\mathbb{P}_n$ denotes the empirical average over a fold in cross validation and $Pred$ is the predicted treatment in the cross validation procedure. The second evaluation criterion was related to significance tests. The test was performed when a 10-fold cross validation procedure is finished, that is, when we had the predicted treatments for all patients. By comparing the predicted treatments with the treatments that were actually assigned to patients, we divided patients into two groups: those who followed the estimated optimal rule and those who did not follow the rule. We tested the difference between the two groups using the two-sample t-test to see whether the group that followed the estimated rule was better than the other group at the significance level $\alpha=0.05$. The whole procedure was repeated 100 times with different fold partitions in the cross validation. We obtained 1000 value functions and 100 p-values. The mean and standard deviation of these value functions, the proportion of significant p-values and median of these p-values are also presented in Table~\ref{tab:epicpaoutcome}. As reference rules, we also considered two fixed treatment rules: assign all patients to (1) cycled therapy arm or (2) culture-based therapy arm. We show in Table~\ref{tab:epicpaoutcome} their value functions from repeated 10-fold cross validation procedures and one-sided p-values from two-sample t-tests. Note that the p-values for fixed rules were not changed during the cross validation procedure.
$\ell_1$-PLS showed the best performance, and identified that one covariate, baseline \textit{Pa} status, is important in estimating ITRs. The significance tests also confirmed its superior performance. Both OWL methods failed to detect the heterogeneity of treatment effects. They assigned all patients to the cycled therapy arm. Although the performances of RWL methods (without variable selection) were only slightly better, on average, than those of OWL methods, they did detect some heterogeneity across treatments. Variable selection further improved the performance in terms of achieving higher values and better significance tests. Moreover, both RWL-VS-Linear and RWL-VS-Gaussian selected the same covariate as $\ell_1$-PLS.

Our results are consistent with results from the trial suggesting that the subgroup of patients who were \textit{Pa} negative at the baseline visit (albeit positive within 6 months of baseline to meet eligibility) may have more greatly benefited from cycled therapy to suppress \textit{Pa} positivity during the trial. However, it is important to note clinically that further analyses of the trial data demonstrated the comparable effectiveness using culture-based therapy as compared to cycled therapy in reducing the overall prevalence of \textit{Pa} positive cultures at the end of the trial.

\begin{table}[tbp]
\centering
\small
\caption{Comparison of methods for estimating ITRs on the EPIC data with the \textit{Pa} endpoint. Higher values are better.}
\label{tab:epicpaoutcome}
\begin{tabular}{lcccc}
\addlinespace
\toprule
& Predicted treatment&  Mean value & Proportion of        &  Median  \\
& \#cycled vs \#culture &  (std)      & significant p-values &  p-value  \\
\midrule
$\ell_1$-PLS & 168 : 115 &  0.913 (0.040) & 1 & 0.004 \\
OWL-Linear & 283 : 0\phantom{00} & 0.894 (0.054) & 0 & 0.111 \\
OWL-Gaussian & 283 : 0\phantom{00} & 0.894 (0.054) & 0 & 0.111 \\
RWL-Linear & 178 : 105 & 0.896 (0.046) & 0.37 & 0.072 \\
RWL-Gaussian & 169 : 114 & 0.900 (0.045) & 0.64 & 0.0404\\
RWL-VS-Linear & 168 : 115 & 0.909 (0.044) & 0.90 & 0.002 \\
RWL-VS-Gaussian & 168 : 115 & 0.907 (0.045) & 0.85 & 0.007\\
\hdashline
Fixed rule (cycled) & 283 : 0\phantom{00} & 0.894 (0.054) & 0 &  0.111 \\
Fixed rule (culture-based) & \phantom{00}0 : 283 & 0.864 (0.054) & 0 & 0.889\\
\bottomrule
\end{tabular}
\end{table}

The number of PEs during the study was a rate variable. We were seeking an ITR to lower the number of PEs, so fewer PEs were desirable. We compared several methods: PoissonReg, $\ell_1$-PoissonReg, OWL-Linear, OWL-Gaussian, RWL-Linear, RWL-Gaussian, RWL-VS-Linear, and RWL-VS-Gaussian. PoissonReg fitted a Poisson regression model using the basis function set $(1,\bm{X},A,\bm{X}A)$, computed the predicted number of PEs for a particular patient at each arm, and assigned the treatment with smaller prediction to this patient. $\ell_1$-PoissonReg was similar with PoissonReg, but used an $\ell_1$ penalized term to perform variable selection. For OWL methods, we used the opposite of individual annual PE rate as the continuous outcome. The annual PE rate for the $i$-th patient is defined as $\frac{r_i}{t_i}*365.25$, where $r_i$ is the number of PEs for the $i$-th patient, and $t_i$ is the duration (in days) when the $i$-th patient stayed in the study. For RWL methods, we used a main effects Poisson regression model to estimate residuals, as described in Section \ref{sec:generalframework}. A 10-fold cross-validation was used to tune parameters. The predicted treatment allocation is shown in Table~\ref{tab:epicpeoutcome}.
A similar evaluation procedure as that for the \textit{Pa} endpoint was also performed, where we used instead the annual PE rate for the group that followed the rule as the first evaluation criterion. The group-wise annual PE rate is given as $\frac{\mathbb{P}_n[R\mathbb{I}(A=Pred)]}{\mathbb{P}_n[T\mathbb{I}(A=Pred)]}*365.25$, where $R$ is the number of PEs, $T$ is the duration (in days) in the study, $\mathbb{P}_n$ denotes the empirical average over a fold in cross validation and $Pred$ is the predicted treatment assignment in the cross validation procedure. For the second criterion, a Poisson regression model was used to test whether the group that followed the estimated rule had fewer PEs than the other group that did not follow the estimated rule. The results are presented in Table~\ref{tab:epicpeoutcome}. We also considered two fixed rules as references.
All RWL methods outperformed OWL and Poisson regression methods in terms of achieving better PE rates and better significance test results. $\ell_1$-PoissonReg identified three important covariates, baseline \textit{Pa} status, baseline S. aureus staus and height, for treatment assignment.

For this case, RWL-VS-Linear and RWL-VS-Gaussian did not improve performances comparing with their counterparts without variable selection.
RWL-VS-Linear did not discard any clinical covariates. As its tuned parameter $\lambda_1$ was very small, RWL-VS-Linear gave a similar performance as RWL-Linear. The estimated rule from RWL-Linear was determined by the following decision function $F$,
\begin{eqnarray*}
F &=& 1.43*\mathbb{I}(gender=female) + 0.07*\textrm{age} -0.58*\mathbb{I}(neverpapos=yes) \\
& & - 1.01*\mathbb{I}(pabl='+') -0.17*\mathbb{I}(sabl='+')+ 0.02*\textrm{weight} + 0.01*\textrm{height}\\ & & -0.17*\mathbb{I}(anyabxhist=yes)+0.11*\textrm{bmi}-0.90*\mathbb{I}(genotype=homozygous)\\
& &-0.83*\mathbb{I}(genotype=heterozygous)-2.90
\end{eqnarray*}
The corresponding ITR is that if $F>0$, assign cycled therapy to this patient, otherwise assign the culture-based therapy.
The proposed rule suggests that female gender and increasing age are key determinants for the potential of a patient to benefit from cycled therapy, which is not surprising since these characteristics are known risk factors for exacerbations.  Counterintuitively, this rule suggests that Pa positivity at baseline leads against the recommendation for cycled therapy unlike with the prior outcome of increased Pa frequency. The rule also suggests that patients with genotypes other than delF508 heterozygous and homozygous may benefit more from cycled therapy as compared to patient with the delF508 mutation. Overall, these results indicate the importance of evaluating multiple endpoints to evaluate consistency in recommendations for treatment based on these proposed rules.

For RWL-VS-Gaussian, three clinical covariates (age, weight, baseline S. aureus status) were identified to be unimportant for estimating ITRs. It is well known that the shrinkage by the $\ell_1$ penalty causes the estimates of the non-zero coefficients to be biased towards zero \citep{Hastie2001}. For similar reasons, the estimates from RWL-VS-Gaussian might be biased. One approach for reducing this bias is to run the $\ell_1$ penalty method to identify important covariates, and then fit a model without the $\ell_1$ term to the selected set of covariates. We applied RWL-Gaussian to the identified set of covariates. The results are also presented in Table~\ref{tab:epicpeoutcome}. The obtained annual PE rate was better than that from RWL-Gaussian alone. Thus variable selection is still helpful to improve the performance.

\begin{table}[tbp]
\centering
\small
\caption{Comparison of methods for estimating ITR on the EPIC data with the PE endpoint. Lower annual PE rates are better.}
\label{tab:epicpeoutcome}
\begin{tabular}{lcccc}
\addlinespace
\toprule
& Predicted treatment   &  Mean annual & Proportion of  &  Median  \\
& \#cycled vs \#culture & PE rate (std) & significant p-values &  p-value  \\
\midrule
PoissonReg & 150 : 133 &  0.742 (0.224) & 0.21 & 0.245 \\
$\ell_1$-PoissonReg & 120 : 163 &  0.707 (0.233) & 0.55 & 0.042 \\
OWL-Linear & 141 : 142 & 0.722 (0.236) & 0.27 & 0.123 \\
OWL-Gaussian & 140 : 143 & 0.714 (0.247) & 0.33 & 0.079 \\
RWL-Linear & 105 : 178 & 0.673 (0.226) & 0.82 & 0.013 \\
RWL-Gaussian & 108 : 175 & 0.667 (0.220) & 0.90 & 0.008 \\
RWL-VS-Linear & 105 : 178 & 0.676 (0.227) & 0.75 & 0.013 \\
RWL-VS-Gaussian & 108 : 175 &  0.688 (0.218) & 0.69 & 0.024 \\
\hdashline
RWL-VS-Gaussian & & & & \\
+ RWL-Gaussian & 102 : 181 & 0.639 (0.220) & 0.97 & 0.001 \\
\hdashline
Fixed rule (cycled) & 283 : 0\phantom{00} & 0.725 (0.245) & 0 &  0.140 \\
Fixed rule (culture-based) & \phantom{00}0 : 283  & 0.823 (0.253) & 0 & 0.860\\
\bottomrule
\end{tabular}
\end{table}

In summary, the proposed RWL methods can identify potentially useful ITRs. For example, the estimated ITR for the \textit{Pa} endpoint in the EPIC trial is that if the baseline \textit{Pa} status is positive, assign culture-based therapy to the patient; if the baseline \textit{Pa} status is negative, assign the cycled therapy.
As the \textit{Pa} endpoint was a secondary endpoint in the trial, the clinical utility of these findings must be weighed in context of all study findings and safety. But nonetheless identifying such a strategy is important to improve personalized clinical practice if the strategy is confirmed by future comparative studies.

\section{Discussion}
In this article, we have proposed a general framework called Residual Weighted Learning (RWL) to use outcome residuals to estimate optimal ITRs. The residuals may be obtained from linear models or generalized linear models, and hence RWL can handle various types of outcomes, including continuous, count and binary outcomes.

The proposed RWL methods appear to achieve better performance compared to other existing methods as shown in the simulation studies and real data analyses. The goal of RWL is to improve finite sample performance of OWL methods. It is not surprising that RWL outperforms OWL in terms of achieving higher value function and smaller variance. Two-step methods, such as $\ell_1$-PLS and $\ell_1$-PoissonReg, are generally parametric. Their performance largely depends on how close the posited model is to the true model of the conditional mean outcome. When the model is correctly specified, two-step methods can be more efficient than nonparametric RWL methods. However, the relationship among clinical covariates, treatment assignment, and outcome is complex in practice. Misspecified models may lead to biased estimates.

Both two-step methods and RWL require correct model specification to guarantee their performance. We consider two levels of model specification. The first one is the model for the decision boundary; the other is the model for the outcome. Two-step methods, such as $\ell_1$-PLS, only specify models for conditional outcomes, and the decision boundary is implied from these models. Thus the performance of two-step methods are only related to these models. However, for RWL, the two levels are separated. RWL methods target the decision boundary directly. The model for decision boundary is more critical to performance.
RWL with Gaussian RBF kernel can approximate any form of decision boundary when the sample size is sufficiently large. Even when using linear kernels, RWL is more robust to model misspecification on the decision boundary than the two-step methods. If the conditional mean outcome is assumed to be linear in the parameters, the decision boundary will be also; the converse does not hold. For RWL with linear kernel, the decision boundary remains a simple linear form, while the conditional mean outcome is allowed to be nonlinear.

Another source of robustness of RWL comes from the model for the outcome, in which we estimate residuals. By Theorem \ref{thm:consistency}, no matter how bad the residual estimate is, RWL is still consistent. In the simulation studies, we have investigated the robustness of residual estimates on finite samples. We applied the null model, as an example of incorrect model specification, to estimate residuals, and achieved satisfactory performance. The simulation studies confirm the robustness of RWL. In this article, we used simple models to obtain the residuals. However, RWL is quite flexible in the methods used to estimate residuals. When the sample size is small or the heterogeneous treatment effect is weak, we may consider a complicated method to accurately estimate residuals to improve the performance of RWL. Alternatively, nonparametric regression methods, such as support vector regression \citep{Vapnik:svm98} or random forests \citep{Breiman:RF2001}, are an option to eliminate model misspecification.

RWL is closely related to a robust augmented inverse probability weighted estimator (AIPWE) of the value function. AIPWE was first introduced into optimal treatment regimes by \citet{Zhang:RobustITR2012}. Under the setup in this article, \textit{i.e.}, $\mathcal{A}=\{1,-1\}$ and $\pi(a,\bm{x})$ is known, for any rule $d$, its associated value function can be estimated by an AIPWE,
\begin{equation*}
\textrm{AIPWE}(d) = \frac{1}{n}\sum_{i=1}^n\left\{\frac{\mathbb{I}(d(\bm{x}_i)=a_i)}{\pi(a_i,\bm{x}_i)}R_i-\frac{\mathbb{I}(d(\bm{x}_i)=a_i)-\pi(a_i,\bm{x}_i)}{\pi(a_i,\bm{x}_i)}\hat{m}(\bm{x}_i)\right\},
\end{equation*}
where $\hat{m}(\bm{x})=\hat{\mu}_1(\bm{x})\mathbb{I}(d(\bm{x})=1)+\hat{\mu}_{-1}(\bm{x})\mathbb{I}(d(\bm{x})=-1)$, $\hat{\mu}_1(\bm{x})$ is an estimate of $\mathbb{E}(R|\bm{X}=\bm{x},A=1)$, and $\hat{\mu}_{-1}(\bm{x})$ is an estimate of $\mathbb{E}(R|\bm{X}=\bm{x},A=-1)$.
If we take $\hat{g}^{\ast}(\bm{x})=\hat{m}(\bm{x})$, for  a fixed $d(\bm{x})$, then maximizing AIPWE is equivalent to minimizing the risk in (\ref{eq:empiriclRWL}). Hence AIPWE is essentially a ``contrast'' weighted learning method, where the contrast is a modelled form that links closely to the residual. Conversely, if we take $\hat{\mu}_1(\bm{x})=\hat{\mu}_{-1}(\bm{x})=\hat{g}^{\ast}(\bm{x})$, where $\hat{g}^{\ast}(\bm{x})$ estimates $g^{\ast}(\bm{x})$ in (\ref{eq:gstar}), then maximizing AIPWE is again equivalent to minimizing (\ref{eq:empiriclRWL}). However, RWL can now be obtained by replacing the 0-1 loss with the smoothed ramp loss, and hence RWL can be viewed as a modification of AIPWE with alternative useful connections to classification methods.

Variable selection is critical for good performance of RWL, as demonstrated in the simulation studies and data analyses. We suggest that practical use of RWL should necessarily be accompanied by variable selection. In Section \ref{sec:rwlvs}, we introduce the concepts of predictive and prescriptive covariates. It has been noticed in practical settings that the main effects, rather than treatment-covariate interactions, tend to explain most of the variability in the outcome, and thus they are important for good prediction \citep{Gunter2011:Qualitative}. It is not uncommon that interactions between prescriptive covariates and treatment are overlooked due to their small predictive ability. Hence variable selection techniques designed for prediction applications might not work perfectly in finding the prescriptive covariates. In contrast, RWL focuses on treatment-covariate interactions, and variable selection in RWL only targets the prescriptive covariates.

RWL methods with linear and Gaussian RBF kernels were discussed in this article. The decision rule from linear RWL is easy to interpret, but at the risk of misspecification.
As a universal kernel, the Gausian RBF kernel is free of model misspecification.
An obvious downside is that the final decision rule may be difficult to interpret, and clinical practitioners may not be comfortable using ``black box'' decision rules. One compromise is to apply linear RWL with a rich set of bases, for example, including two-way and three-way covariate interactions.

When the endpoint outcome is the survival time, observations are commonly subject to right censoring because of subject dropout or censoring due to the end of study. Currently, RWL does not cover this type of outcomes in theory due to the censorship. We recommend using martingale residuals \citep{Fleming:Counting1991} as alternative outcomes. Specifically, residuals in RWL could be opposites of martingale residuals that are estimated by a Cox proportional hazards model on clinical covariates but excluding treatment assignment. Actually, the idea is not new. \citet{Therneau:MResid1990} suggested that martingale residuals from a null Cox model could be used as outcomes for Classification and Regression Trees (CART).
The rigourous theoretical justification is under development. Here we provide an intuitive explanation. Let $T$ denote the survival time, and $C$ denote the censoring time. The observed data quadruplet is $(\bm{X}, A, R=T\wedge C, \Delta=\mathbb{I}(T\leq C))$. Firstly, for general survival analysis, $(R, \Delta)$ is seen as the censored outcome. We view the survival data from a different perspective. We treat $\Delta$ as the outcome, and $(\bm{X}, R)$ as clinical covariates.
In the framework of counting processes, the martingale residual can be explained as the difference between the observed number of failures (0 or
1) during the time in the study for a subject and the expected numbers based on the fitted model. Thus it is almost in line with our derivation of RWL, although $R$ may not be appropriate as a clinical covariate. Secondly, martingale residuals from a Cox model fitted excluding a covariate of interest can be used to determine the appropriate functional form of the covariate \citep{Therneau:MResid1990}. If we estimate martingale residuals by ignoring treatment assignment, they contain information related to the heterogeneity of treatment effects. Thirdly, martingale residuals have some properties reminiscent of linear models, for example, the residuals are asymptotically uncorrelated. Moreover, the weighted sum of martingale residuals is zero, which is essential for RWL to control the treatment matching factor.

Several extensions are worthy of further investigation. In this article, we only considered two-arm trials. In practice, some trials involve multiple arms, \textit{i.e.} have three or more treatment arms. A recent review by \citet{Baron2013:multiarm} reported that $17.6\%$ of published randomized clinical trials in 2009 were multiple-arm. Because of the number of possible comparisons, it is more complex to find ITRs in such trials compared with two-arm trials. It would thus be worthwhile to extend RWL to multiple-arm trials. In the machine learning literature, two schools of approaches are generally suggested for multi-category classification. One is to directly take all classes into consideration. Multicategory SVMs \citep{Lee2004:MSVM} and penalized logistic regression \citep{Zhu2004:PLR} are two examples. The other school is to construct and combine several binary classifiers. One-versus-one, one-versus-all, and more general error-correcting output codes \citep[ECOC]{Dietterich1995:ECOC} are in this school. Similar generalizations may be possible for finding ITRs for multiple-arm trials.

For chronic diseases, multi-stage dynamic treatment regimes are more useful than single-stage optimal treatment regimes. In the context of multi-stage decision problems, a dynamic treatment regime is a sequence of decision rules, one per stage of intervention, for adapting a treatment plan over time to an individual. \citet{Zhao2014:DTR} has extended OWL to dynamic treatment regimes. Considering the improved performance of RWL over OWL, it is of interest to extend RWL to dynamic treatment regimes.

The intended use of the estimated optimal ITRs discovered using the proposed approach is for treating new patients. The theoretical, simulation and data analysis results demonstrate that such estimated ITRs are likely to lead to improved clinical outcomes for these patients. However, since the data used for estimating the ITRs should not also be used for confirmation, it is expected that an additional randomized clinical trial comparing the estimated ITR with standard of care, or some other suitable treatment comparison, will be used for confirmation as is typically done for new candidate treatments \citep{Zhao2011:Reinforcement}.



\appendix
\makeatletter   
 \renewcommand{\@seccntformat}[1]{APPENDIX~{\csname the#1\endcsname}.\hspace*{1em}}
 \makeatother

\section{Proofs}
\textbf{Proof of Lemma 2.1}
\begin{proof}Note that,
\begin{equation} \label{eq:factor}
\mathbb{E}\left(\frac{\mathbb{I}\big(A\neq d(\bm{X})\big)}{\pi(A,\bm{X})}\Big|\bm{X}\right)=\mathbb{E}\left(\mathbb{I}\big(d(\bm{X})\neq1\big)|\bm{X}, A=1\right)+\mathbb{E}\left(\mathbb{I}\big(d(\bm{X})\neq-1\big)|\bm{X}, A=-1\right)=1.
\end{equation}
The desired result follows easily.
\end{proof}

\textbf{Proof of Theorem 2.2}
\begin{proof} For any measurable function $g$,
\begin{eqnarray*}
&&Var\left(\frac{R-g(\bm{X})}{\pi(A,\bm{X})}\mathbb{I}\big(A\neq d(\bm{X})\big)\right)\\
&=&Var\left(\frac{R-\tilde{g}(\bm{X})}{\pi(A,\bm{X})}\mathbb{I}\big(A\neq d(\bm{X})\big)\right)+Var\left(\frac{\tilde{g}(\bm{X})-g(\bm{X})}{\pi(A,\bm{X})}\mathbb{I}\big(A\neq d(\bm{X})\big)\right) \\
& &+2Cov\left(\frac{R-\tilde{g}(\bm{X})}{\pi(A,\bm{X})}\mathbb{I}\big(A\neq d(\bm{X})\big),\frac{\tilde{g}(\bm{X})-g(\bm{X})}{\pi(A,\bm{X})}\mathbb{I}\big(A\neq d(\bm{X})\big)\right).
\end{eqnarray*}
It suffices to show that the covariance term is zero. Applying (\ref{eq:factor}), we have
\begin{eqnarray*}
&&\mathbb{E}\left(\frac{R-\tilde{g}(\bm{X})}{\pi(A,\bm{X})}\mathbb{I}\big(A\neq d(\bm{X})\big)\Big|\bm{X}\right) \\
&=&\mathbb{E}\left(\frac{R}{\pi(A,\bm{X})}\mathbb{I}\big(A\neq d(\bm{X})\big)\Big|\bm{X}\right)-\tilde{g}(\bm{X})\mathbb{E}\left(\frac{\mathbb{I}\big(A\neq d(\bm{X})\big)}{\pi(A,\bm{X})}\Big|\bm{X}\right)\\
&=&0.
\end{eqnarray*}
Note that
\begin{eqnarray} 
\tilde{g}(\bm{X})&=&\mathbb{E}\left(\frac{R}{\pi(A,\bm{X})}\mathbb{I}\big(A\neq d(\bm{X})\big)\Big|\bm{X}\right)\nonumber \\
&=&\mathbb{E}(R|\bm{X},A=1)\mathbb{I}\big(d(\bm{X})\neq1\big)+\mathbb{E}(R|\bm{X},A=-1)\mathbb{I}\big(d(\bm{X})\neq-1\big). \label{eq:tildeg}
\end{eqnarray}
Thus we have,
\begin{eqnarray*}
&&\mathbb{E}\left(\frac{R-\tilde{g}(\bm{X})}{\pi(A,\bm{X})^2}\mathbb{I}\big(A\neq d(\bm{X})\big)\Big|\bm{X}\right)\\
&=&\mathbb{E}\left(R-\tilde{g}(\bm{X})|\bm{X},A=1\right)\mathbb{I}\big(d(\bm{X})\neq1\big)/\pi(1,\bm{X})\\
& & +\mathbb{E}\left(R-\tilde{g}(\bm{X})|\bm{X},A=-1\right)\mathbb{I}\big(d(\bm{X})\neq-1\big)/\pi(-1,\bm{X})\\
&=&0.
\end{eqnarray*}
The desired result follows easily.
\end{proof}

\textbf{Proof of Theorem 3.1}
\begin{proof}
Given $\bm{X}=\bm{x}$, for any measurable function $f$, similar reasoning to that used in the proof of Lemma 2.1 yields,
\begin{equation*}
\mathbb{E}\left(\frac{T\big(Af(\bm{X})\big)}{\pi(A,\bm{X})}\Big|\bm{X}=\bm{x}\right)=2.
\end{equation*}
Then the conditional $T$-risk is
\begin{eqnarray}
&&\mathbb{E}\left(\frac{R-g(\bm{X})}{\pi(A,\bm{X})}T(Af(\bm{X}))\big|\bm{X}=\bm{x}\right)\nonumber \\
&=&\mathbb{E}(R|\bm{X}=\bm{x},A=1)T(f(\bm{x}))+\mathbb{E}(R|\bm{X}=\bm{x},A=-1)T(-f(\bm{x})) - 2g(\bm{x})\nonumber\\
&=&\left(\mathbb{E}(R|\bm{X}=\bm{x},A=1)-\mathbb{E}(R|\bm{X}=\bm{x},A=-1)\right)T(f(\bm{x})) \nonumber\\
&& +2\mathbb{E}(R|\bm{X}=\bm{x},A=-1)- 2g(\bm{x}). \label{eq:triskf}
\end{eqnarray}
If $\mathbb{E}\left[R|\bm{X}=\bm{x},A=1\right]-\mathbb{E}\left[R|\bm{X}=\bm{x},A=-1\right]>0$, any function $f(\bm{x})\geq 1$ minimizes the conditional $T$-risk; similarly, if $\mathbb{E}\left[R|\bm{X}=\bm{x},A=1\right]-\mathbb{E}\left[R|\bm{X}=\bm{x},A=-1\right]<0$, any function
$f(\bm{x})\leq -1$ minimizes the conditional $T$-risk. For either case, $sign({f}^*_{T,g})=d^*$.

For the second part, by applying (\ref{eq:triskf}),
\begin{eqnarray*}
&&\mathbb{E}\left(\frac{R-g(\bm{X})}{\pi(A,\bm{X})}T(Ad^*(\bm{X}))|\bm{X}=\bm{x}\right)-\mathbb{E}\left(\frac{R-g(\bm{X})}{\pi(A,\bm{X})}T(Af_{T,g}^*(\bm{X}))|\bm{X}=\bm{x}\right)\\
&=& \left(\mathbb{E}(R|\bm{X}=\bm{x},A=1)-\mathbb{E}(R|\bm{X}=\bm{x},A=-1)\right)\left(T(d^*(\bm{x}))-T(f_{T,g}^*(\bm{x})) \right) = 0.
\end{eqnarray*}
The desired result follows by taking expectations on both sides.
\end{proof}

\textbf{Proof of Theorem 3.2}
\begin{proof}
Given $\bm{X}=\bm{x}$. By applying (\ref{eq:triskf}), for any measurable function $f$, we have
\begin{eqnarray*}
&&\mathbb{E}\left(\frac{R-g(\bm{X})}{\pi(A,\bm{X})}T(Af(\bm{X}))|\bm{X}=\bm{x}\right)-\mathbb{E}\left(\frac{R-g(\bm{X})}{\pi(A,\bm{X})}T(Af_{T,g}^*(\bm{X}))|\bm{X}=\bm{x}\right)\\
&=& \left(\mathbb{E}(R|\bm{X}=\bm{x},A=1)-\mathbb{E}(R|\bm{X}=\bm{x},A=-1)\right)\left(T(f(\bm{x}))-T(f_{T,g}^*(\bm{x})) \right).
\end{eqnarray*}
Similarly,
\begin{eqnarray*}
&&\mathbb{E}\left(\frac{R}{\pi(A,\bm{X})}\mathbb{I}(A\neq sign(f(\bm{X})))|\bm{X}=\bm{x}\right) - \mathbb{E}\left(\frac{R}{\pi(A,\bm{X})}\mathbb{I}(A\neq d^*(\bm{X}))|\bm{X}=\bm{x}\right)\\
&=& \left(\mathbb{E}(R|\bm{X}=\bm{x},A=1)-\mathbb{E}(R|\bm{X}=\bm{x},A=-1)\right)\left(\mathbb{I}(sign(f(\bm{x}))\neq 1)-\mathbb{I}(d^*(\bm{x})\neq 1)\right).
\end{eqnarray*}
From the proof of Theorem 3.1, when $\mathbb{E}(R|\bm{X}=\bm{x},A=1)>\mathbb{E}(R|\bm{X}=\bm{x},A=-1)$, $f_{T,g}^*(\bm{x})\geq 1$ and $d^*(\bm{x})=1$, so $T(f_{T,g}^*(\bm{x}))=0$ and $\mathbb{I}(d^*(\bm{x})\neq 1)=0$; when $\mathbb{E}(R|\bm{X}=\bm{x},A=1)<\mathbb{E}(R|\bm{X}=\bm{x},A=-1)$, $f_{T,g}^*(\bm{x})\leq -1$
and $d^*(\bm{x})=-1$, so $T(f_{T,g}^*(\bm{x}))=2$ and $\mathbb{I}(d^*(\bm{x})\neq 1)=1$.
Note that, for any measurable function $f$, $1\geq T(f(\bm{x})) - \mathbb{I}(sign(f(\bm{x}))\neq 1) \geq0$. Thus it is easy to check that when $\mathbb{E}(R|\bm{X}=\bm{x},A=1)>\mathbb{E}(R|\bm{X}=\bm{x},A=-1)$, $$T(f(\bm{x}))-T(f_{T,g}^*(\bm{x}))\geq \mathbb{I}(sign(f(\bm{x}))\neq 1)-\mathbb{I}(d^*(\bm{x})\neq 1),$$ and when $\mathbb{E}(R|\bm{X}=\bm{x},A=1)<\mathbb{E}(R|\bm{X}=\bm{x},A=-1)$, $$T(f(\bm{x}))-T(f_{T,g}^*(\bm{x}))\leq \mathbb{I}(sign(f(\bm{x}))\neq 1)-\mathbb{I}(d^*(\bm{x})\neq 1).$$
So, for either case, we have
\begin{eqnarray*}
& &\mathbb{E}\left(\frac{R}{\pi(A,\bm{X})}\mathbb{I}(A\neq sign(f(\bm{X})))|\bm{X}=\bm{x}\right) - \mathbb{E}\left(\frac{R}{\pi(A,\bm{X})}\mathbb{I}(A\neq d^*(\bm{X}))|\bm{X}=\bm{x}\right) \\
&\leq &
\mathbb{E}\left(\frac{R-g(\bm{X})}{\pi(A,\bm{X})}T(Af(\bm{X}))|\bm{X}=\bm{x}\right)-\mathbb{E}\left(\frac{R-g(\bm{X})}{\pi(A,\bm{X})}T(Af_{T,g}^*(\bm{X}))|\bm{X}=\bm{x}\right).
\end{eqnarray*}
The desired result follows by taking expectations on both sides.
\end{proof}

\textbf{Proof of Theorem 3.3}
\begin{proof}
Let $L(h,b)=(R-g(\bm{X}))T(A(h(\bm{X})+b))/\pi(A,\bm{X})$. For simplicity, we denote $f_{D_n,\lambda_n}$, $h_{D_n,\lambda_n}$ and $b_{D_n,\lambda_n}$ by $f_n$, $h_n$ and $b_n$, respectively. By the definition of $h_{D_n,\lambda_n}$ and $b_{D_n,\lambda_n}$, we have, for any $h\in\mathcal{H}_K$ and $b\in\mathbb{R}$,
\begin{equation*}
\mathbb{P}_n(L(h_n,b_n))\leq\mathbb{P}_n(L(h_n,b_n))+\frac{\lambda_n}{2}||h_n||^2_K \leq \mathbb{P}_n(L(h,b))+\frac{\lambda_n}{2}||h||^2_K,
\end{equation*}
where $\mathbb{P}_n$ denotes the empirical measure of the observed data. Then,
$\lim\sup_n\mathbb{P}_n(L(h_n,b_n))\leq\mathbb{P}(L(h,b))=\mathcal{R}_{T,g}(h+b)$ with probability 1. This implies
\begin{equation*}
\lim\sup_n\mathbb{P}_n(L(h_n,b_n))\leq\inf_{h\in\mathcal{H}_K, b\in\mathbb{R}}\mathcal{R}_{T,g}(h+b)\leq \mathbb{P}(L(h_n,b_n))
\end{equation*}
with probability 1. It suffices to show $\mathbb{P}_n(L(h_n,b_n))-\mathbb{P}(L(h_n,b_n))\rightarrow0$ in probability.

We first obtain a bound for $||h_n||_K$. Since $\mathbb{P}_n(L(h_n,b_n))+\lambda_n||h_n||^2_K/2\leq\mathbb{P}_n(L(h,b))+\lambda_n||h||^2_K/2$, for any $h\in \mathcal{H}_K$ and $b\in\mathbb{R}$, we can choose  $h=0$ and $b=0$ to obtain,
$\mathbb{P}_n(L(h_n,b_n))+\lambda_n||h_n||^2_K/2\leq\mathbb{P}_n((R-g(\bm{X}))/\pi(A,\bm{X}))$. Note that $0\leq T(u) \leq 2$. We thus have,
\begin{equation*}
\lambda_n||h_n||^2_K \leq 2\mathbb{P}_n(|R-g(\bm{X})|/\pi(A,\bm{X}))\leq 2M_0.
\end{equation*}
Let $M_1=\sqrt{2M_0}$. Then the $\mathcal{H}_K$ norm of $\sqrt{\lambda_n}h_n$ is bounded by $M_1$.

Next we obtain a bound for $b_n$. We claim that there is a global solution $(h_n,b_n)$ such that $h_n(\bm{x}_i)+b_n\in[-1,1]$ for some $i$. Suppose there is a global solution $(h'_n,b'_n)$ such that $|h'_n(\bm{x}_i)+b'_n|>1$ for all $i$. Let $\delta=|h'_n(\bm{x}_{i_0})+b'_n|=\min_{1\leq i\leq n}|h'_n(\bm{x}_{i})+b'_n|>1$. Then let $h_n=h'_n$ and $b_n=b'_n-(\delta-1)sign(h'_n(\bm{x}_{i_0})+b'_n)$. It is easy to check that $h_n(\bm{x}_{i_0})+b_n=1$ if $h'_n(\bm{x}_{i_0})+b'_n>1$, and  $h_n(\bm{x}_{i_0})+b_n=-1$ if $h'_n(\bm{x}_{i_0})+b'_n<-1$; furthermore when $i\neq i_0$, $h_n(\bm{x}_{i})+b_n\geq 1$ if $h'_n(\bm{x}_{i})+b'_n> 1$, and $h_n(\bm{x}_{i})+b_n\leq -1$ if $h'_n(\bm{x}_{i})+b'_n< -1$. So $T(h_n(\bm{x}_{i})+b_n)=T(h'_n(\bm{x}_{i})+b'_n)$ for all $i$. Hence $(h_n,b_n)$ is a global solution and satisfies our claim. Now if a solution $(h_n,b_n)$ satisfies our claim, we then have,
\begin{equation*}
|b_n|\leq 1+|h_n(\bm{x}_{i_0})|\leq 1+||h_n||_{\infty}.
\end{equation*}
Note that $||h||_{\infty}\leq C_K||h||_K$. We have,
\begin{equation*}
|\sqrt{\lambda_n}b_n|\leq \sqrt{\lambda_n}+C_K\sqrt{\lambda_n}||h_n||_K.
\end{equation*}
Since $\lambda_n\rightarrow0$, and $C_K$ and $\sqrt{\lambda_n}||h_n||_K$ are both bounded, we have $|\sqrt{\lambda_n}b_n|$ is bounded too. Let the bound be $M_2$, \textit{i.e.} $|\sqrt{\lambda_n}b_n|\leq M_2$.

Note that the class $\{\sqrt{\lambda_n}h:||\sqrt{\lambda_n}h||_K\leq M_1\}$ is a Donsker class. So $\{\sqrt{\lambda_n}(h+b):||\sqrt{\lambda_n}h||_K\leq M_1, |\sqrt{\lambda_n}b|\leq M_2\}$ is also P-Donsker. Consider the function
\begin{equation*}
T_{\lambda}(u)=\begin{cases}
2\sqrt{\lambda} & \text{if } \quad u < -\sqrt{\lambda},\\
2\sqrt{\lambda} - \frac{1}{\sqrt{\lambda}}(\sqrt{\lambda} + u)^2 & \text{if } \quad -\sqrt{\lambda} \leq u < 0 ,\\
\frac{1}{\sqrt{\lambda}}(\sqrt{\lambda} - u)^2 & \text{if } \quad \quad 0 \leq u < \sqrt{\lambda},\\
0 & \text{if } \quad u \geq \sqrt{\lambda}.
\end{cases}
\end{equation*}
We have $T_{\lambda}(\sqrt{\lambda}u)=\sqrt{\lambda}T(u)$. Since $T_{\lambda}(u)$ is a Lipschitz continuous function with Lipschitz constant equal to 2, and $\frac{R-g(\bm{X})}{\pi(A,\bm{X})}$ is bounded, the class $\{\sqrt{\lambda_n}L(h,b):||\sqrt{\lambda_n}h||_K\leq M_1, |\sqrt{\lambda_n}b|\leq M_2\}$ is also P-Donsker. Therefore,
\begin{equation*}
\sqrt{n\lambda_n}(\mathbb{P}_n-\mathbb{P})L(h_n, b_n)=O_p(1).
\end{equation*}
Consequently, from $n\lambda_n\rightarrow\infty$, $\mathbb{P}_n(L(h_n,b_n))-\mathbb{P}(L(h_n,b_n))\rightarrow 0$ in probability.
\end{proof}

\textbf{Proof of Lemma 3.4}
\begin{proof} Fix any $0<\epsilon<1$.
$d^*(\bm{x})=sign(\mathbb{E}(R|\bm{X}=\bm{x},A=1)-\mathbb{E}(R|\bm{X}=\bm{x},A=-1))$ is measurable. Since $\mu$ is regular, using Lusin's theorem in measure theory, we know that $d^*(\bm{x})$ can be approximated by a continuous function $f'(\bm{x})\in C(\mathcal{X})$ such that $\mu(f'(\bm{x})\neq d^*(\bm{x}))\leq \frac{\epsilon}{4M}$. Thus

\begin{eqnarray*}
&&\mathbb{E}\left(\frac{R-g(\bm{X})}{\pi(A,\bm{X})}T(Af'(\bm{X}))|\bm{X}=\bm{x}\right)-\mathbb{E}\left(\frac{R-g(\bm{X})}{\pi(A,\bm{X})}T(Ad^*(\bm{X}))|\bm{X}=\bm{x}\right)\\
&=& \left(\mathbb{E}(R|\bm{X}=\bm{x},A=1)-\mathbb{E}(R|\bm{X}=\bm{x},A=-1)\right)\left(T(f'(\bm{x}))-T(d^*(\bm{x})) \right).
\end{eqnarray*}
Then,
\begin{eqnarray*}
&&\mathcal{R}_{T,g}(f')-\mathcal{R}_{T,g}^* = |\mathcal{R}_{T,g}(f')-\mathcal{R}_{T,g}(d^*)|\\
&=&\Big|\int\left(\mathbb{E}(R|\bm{X}=\bm{x},A=1)-\mathbb{E}(R|\bm{X}=\bm{x},A=-1)\right)\big(T(f'(\bm{x}))-T(d^*(\bm{x}))\big)\mu(d\bm{x})\Big| \\ &\leq& \int \Big|\left(\mathbb{E}(R|\bm{X}=\bm{x},A=1)-\mathbb{E}(R|\bm{X}=\bm{x},A=-1)\right)\Big|\Big|T(f'(\bm{x}))-T(d^*(\bm{x}))\Big|\\
& & \quad \mathbb{I}(f'(\bm{x})\neq d^*(\bm{x}))\mu(d\bm{x}).
\end{eqnarray*}
Since $|R|\leq M$ and $0\leq T(u)\leq 2$,
\begin{equation*}
\mathcal{R}_{T,g}(f')-\mathcal{R}_{T,g}^*< \epsilon
\end{equation*}
Since $K$ is universal, there exist a function $f''\in\mathcal{H}_K$ such that $||f''-f'||_{\infty}<\frac{\epsilon}{4M}$. Note that $T(\cdot)$ is Lipschitz continuous with Lipschitz constant 2. Similarly,
\begin{eqnarray*}
&&|\mathcal{R}_{T,g}(f'')-\mathcal{R}_{T,g}(f')| \\ &=&\Big|\int\left(\mathbb{E}(R|\bm{X}=\bm{x},A=1)-\mathbb{E}(R|\bm{X}=\bm{x},A=-1)\right)\big(T(f''(\bm{x}))-T(f'(\bm{x}))\big)\mu(d\bm{x})\Big| \\ &\leq& 2\int \Big|\left(\mathbb{E}(R|\bm{X}=\bm{x},A=1)-\mathbb{E}(R|\bm{X}=\bm{x},A=-1)\right)\Big|\Big|f'(\bm{x})-f'(\bm{x})\Big|\mu(d\bm{x}) <\epsilon.
\end{eqnarray*}
By combining the two inequalities, we have
\begin{equation*}
\mathcal{R}_{T,g}(f'')-\mathcal{R}_{T,g}^*<2\epsilon.
\end{equation*}
Noting that $f''\in\mathcal{H}_K$ and letting $\epsilon\rightarrow0$, we obtain the desired result.
\end{proof}

\textbf{Proof of Theorem 3.6}
\begin{proof}
The proof follows the idea in \citet[Theorem 7.2]{Devroye:PatternRecog1996}. Since the proof is very similar, we only provide a sketch to save space.

Let $b=0.b_1b_2b_3\cdots$ be a real number on $[0,1]$ with the given binary expansion, and let $B$ be a random variable uniformly distributed on $[0,1]$ with expansion $B=0.B_1B_2B_3\cdots$. Let us restrict ourselves to a random variable $\bm{X}$ with the support $\{\bm{x}_1,\bm{x}_2,\cdots\}$ where $\bm{x}_i\in\mathcal{X}$. For simplicity, we recode the support of $\bm{X}$ as $\{1,2,\cdots\}$. Let
\begin{equation}
P(\bm{X}=i) = p_i, \quad i\geq1,
\end{equation}
where $p_1\geq p_2\geq\cdots>0$, and $\sum_{i=n+1}^{\infty}p_i\geq \max(8c_n, 32np_{n+1})$ for every $n$. Such $p_i$'s exist by \citet[Lemma 7.1]{Devroye:PatternRecog1996}. Let $A\in\{1,-1\}$ be a binomial variable with $\pi(A,\bm{X})=0.5$. For a given $b$, set $R=AM$ if $b_{\bm{X}}=1$, and $R=-AM$ if $b_{\bm{X}}=0$. Then the Bayes rule is $d^*(\bm{X})=(2*b_{\bm{X}}-1)$. Thus each $b\in[0,1]$ describes a different distribution of $(\bm{X},A,R)$. Introduce the shortened notation $D_n =\{(\bm{X}_1,A_1,R_1), \cdots, (\bm{X}_n,A_n,R_n)\}$. Let $d_n$ be a rule generated by data $D_n$. Define $d^i_{n} = d_n(i)$ for $i=1,\cdots,n$. Let $\Delta\mathcal{R}_n(b)$ be the excess risk of the rule $d_n$ for the distribution parametrized by $b$, and $\Delta\mathcal{R}_n(B)$ be the excess risk of the rule $d_n$ for the random distribution.
\begin{eqnarray*}
& &\Delta\mathcal{R}_n(B) \\
&=& \mathbb{E}\left[\frac{R}{\pi(A,\bm{X})}\mathbb{I}(A\neq d_n(\bm{X}))\Big|B\right] - \mathbb{E}\left[\frac{R}{\pi(A,\bm{X})}\mathbb{I}(A\neq d^*(\bm{X}))\Big|B\right] \\
&=&\mathbb{E}\Big[\big(\mathbb{E}(R|B,\bm{X},A=1)-\mathbb{E}(R|B,\bm{X},A=-1)\big)\big(\mathbb{I}(d_n(\bm{X})\neq1)-\mathbb{I}(d^*(\bm{X})\neq1)\big)\big|B\Big]\\
&=&2M\mathbb{E}\big(\mathbb{I}(d_n(\bm{X})\neq d^*(\bm{X}))\big|B\big) \\
&=&2M\mathbb{E}\big(\mathbb{I}(d_n(\bm{X})\neq 2B_{\bm{X}}-1)\big).
\end{eqnarray*}
Let $L_n(B)=\mathbb{E}\big(\mathbb{I}(d_n(\bm{X})\neq 2B_{\bm{X}}-1)\big)$. Then we have,
\begin{equation*}
L_n(B)=\sum_{i=1}^{\infty}p_i\mathbb{I}(d^i_n\neq 2B_i-1).
\end{equation*}
Following the same arguments used in \citet[Theorem 7.2]{Devroye:PatternRecog1996}, we have
\begin{eqnarray*}
P(L_n(B)<2c_n|D_n)\leq P(\sum_{i=n+1}^{\infty}p_iB_i<2c_n)\leq e^{-2n}.
\end{eqnarray*}
Hence we have
\begin{eqnarray*}
\sup_{b}\inf_{n}\mathbb{E}\left(\frac{L_n(b)}{2c_n}\right) & \geq \ & \mathbb{E}\left(\mathbb{E}\left(\inf_n\left(\frac{L_n(b)}{2c_n}\Big|\bm{X}_1,\bm{X}_2,\cdots\right)\right)\right) \\
& \geq & \mathbb{E}\left(1-\sum_{i=1}^{\infty}\mathbb{E}(P(L_n(B)<2c_n|D_n)|\bm{X}_1,\bm{X}_2,\cdots)\right) \\
&\geq& 1-\sum_{i=1}^{\infty}e^{-2n} = \frac{e^2-2}{e^2-1} >\frac{1}{2}.
\end{eqnarray*}
Here we are omitting many steps. Refer to \citet[Theorem 7.2]{Devroye:PatternRecog1996} for details. The conclusion is that there exists a $b$ for which $\Delta\mathcal{R}_n(b)\geq 2Mc_n$, $n=1,2,\cdots$.
\end{proof}

\textbf{Proof of Theorem 3.7}
\begin{proof}
Define the random variable $S = \frac{R-g(\bm{X})}{\pi(A,\bm{X})}$. We consider a probability measure on the triplet $(\bm{X},A,S)$ instead of on $(\bm{X},A,R)$. Let $D_n=\{\bm{X}_i, A_i, S_i\}_{i=1}^n$ be independent random variables with the same distribution as $(\bm{X},A,S)$. Let $\mathbb{P}_n$ be the empirical measure on $D_n$.
For simplicity, we denote $f_{D_n,\lambda_n}$, $h_{D_n,\lambda_n}$ and $b_{D_n,\lambda_n}$ by $f_n$, $h_n$ and $b_n$, respectively.
Let $(\tilde{h}_{\lambda_n}, \tilde{b}_{\lambda_n})$ be a solution of the following optimization problem:
\begin{equation*}
\min_{h\in \mathcal{H}_K, b\in\mathbb{R}}\frac{\lambda_n}{2}||h||_K^2+\mathcal{R}_{T,g}(h+b).
\end{equation*}

Let $L(h,b)=ST(A(h(\bm{X})+b))$. Then
\begin{eqnarray*}
& & \mathcal{R}_{T,g}(f_n) - \mathcal{R}_{T,g}(f^*_{T,g}) \\
&\leq & (\mathbb{P}-\mathbb{P}_n)L(h_n,b_n) + (\frac{\lambda_n}{2}||h_n||^2+\mathbb{P}_nL(h_n,b_n))-(\frac{\lambda_n}{2}||\tilde{h}_{\lambda_n}||^2+\mathbb{P}_nL(\tilde{h}_{\lambda_n},\tilde{b}_{\lambda_n})) \\
& & + (\mathbb{P}_n-\mathbb{P})L(\tilde{h}_{\lambda_n},\tilde{b}_{\lambda_n}) + \mathcal{A}(\lambda_n) \\
& \leq & (\mathbb{P} -\mathbb{P}_n)L(h_n,b_n) + (\mathbb{P}_n-\mathbb{P})L(\tilde{h}_{\lambda_n},\tilde{b}_{\lambda_n})) + \mathcal{A}(\lambda_n).
\end{eqnarray*}
We first estimate the second term by the Hoeffding inequality \citep[Theorem 6.10]{Steinwart:SVM2008}. Since $|L(h,b)|\leq 2M_0$, we thus have, with probability at least $1-\delta/2$,
\begin{equation} \label{eq:secondterm}
(\mathbb{P}_n-\mathbb{P})L(\tilde{h}_{\lambda_n},\tilde{b}_{\lambda_n})\leq M_0\sqrt{\frac{2\log\frac{2}{\delta}}{n}}.
\end{equation}

By the arguments used in the proof of Theorem 3.3, we have $||h_n||_K \leq \sqrt{\frac{2M_0}{\lambda_n}}$ and $|b_n| \leq  1 + C_K\sqrt{\frac{2M_0}{\lambda_n}}$. Then let $\mathcal{F}=\{(h,b)\in \mathcal{H}_K\times\mathbb{R}: ||h||_K\leq \sqrt{\frac{2M_0}{\lambda_n}}, |b|\leq 1+C_K\sqrt{\frac{2M_0}{\lambda_n}}\}$. Let $\tilde{L}(h,b)=S\Big[T(A(h(\bm{X})+b))-1\Big]$.
For the first term, $(\mathbb{P}-\mathbb{P}_n)L(h_n,b_n)$,
\begin{eqnarray*}
(\mathbb{P}-\mathbb{P}_n)L(h_n,b_n) & \leq & \sup_{(h,b)\in\mathcal{F}}(\mathbb{P}-\mathbb{P}_n)L(h,b) \\
&=&\sup_{(h,b)\in\mathcal{F}}(\mathbb{P}-\mathbb{P}_n)\tilde{L}(h,b)
+ (\mathbb{P}- \mathbb{P}_n)L(0,0).
\end{eqnarray*}
When an $(\bm{x}_i, a_i, s_i)$ triplet changes, the random variable $\sup_{(h,b)\in\mathcal{F}}(\mathbb{P}-\mathbb{P}_n)\tilde{L}(h,b)$ can change by no more than $\frac{2M_0}{n}$. McDiarmid's inequality \citep[Theorem 9]{Bartlett:Rademacher2002} then implies that with probability at least $1-\delta/4$,
\begin{equation*}
\sup_{(h,b)\in\mathcal{F}}(\mathbb{P}-\mathbb{P}_n)\tilde{L}(h,b) \leq \mathbb{E}\sup_{(h,b)\in\mathcal{F}}(\mathbb{P}-\mathbb{P}_n)\tilde{L}(h,b) + M_0\sqrt{\frac{2\log(4/\delta)}{n}}.
\end{equation*}
A similar argument, together with the fact that $\mathbb{E}\mathbb{P}_nL(0,0)=\mathbb{P}L(0,0)$, shows that with probability at least $1-\delta/2$,
\begin{equation*}
(\mathbb{P}-\mathbb{P}_n)L(h_n,b_n) \leq \mathbb{E}\sup_{(h,b)\in\mathcal{F}}(\mathbb{P}-\mathbb{P}_n)\tilde{L}(h,b) + 2M_0\sqrt{\frac{2\log(4/\delta)}{n}}.
\end{equation*}
Let $D'_n=\{\bm{X}'_i, A'_i, S'_i\}_{i=1}^n$ be an independent sample with the same distribution as $(\bm{X},A,S)$. Let $\mathbb{P}'_n$ denote the empirical measure on $D'_n$. Let $\sigma$ be a uniform $\{\pm1\}$-valued random variable, and $\sigma_1,\ldots,\sigma_n$ be $n$ independent copies of $\sigma$.
Then we have
\begin{eqnarray*}
\mathbb{E}\sup_{(h,b)\in\mathcal{F}}(\mathbb{P}-\mathbb{P}_n)\tilde{L}(h,b) & = & \mathbb{E}\sup_{(h,b)\in\mathcal{F}}\mathbb{E}\Big(\mathbb{P}'_n\tilde{L}(h,b) -\mathbb{P}_n\tilde{L}(h,b)\Big|D_n\Big) \\
&\leq & 2\mathbb{E}\sup_{(h,b)\in\mathcal{F}}\mathbb{P}_n\sigma\tilde{L}(h,b) \\
&\leq & 2\mathbb{E}\mathbb{E}\Big(\sup_{(h,b)\in\mathcal{F}}|\mathbb{P}_n\sigma\tilde{L}(h,b)|\Big|S_i,A_i,i=1,\ldots,n\Big) \\
&\leq & \frac{16M_0}{n}\mathbb{E}\sup_{(h,b)\in\mathcal{F}}\Big|\sum_{i=1}^n\sigma_i\big(h(\bm{X}_i)+b\big)\Big|.
\end{eqnarray*}
The last inequality is due to the contraction inequality \citep[Corollary 3.17]{Ledoux:Banach1991}. The preceding can be further majorized by using Lemma 22 in \citet{Bartlett:Rademacher2002},
\begin{eqnarray*}
\mathbb{E}\sup_{(h,b)\in\mathcal{F}}(\mathbb{P}-\mathbb{P}_n)\tilde{L}(h,b) & \leq & \frac{16M_0}{n}\mathbb{E}\sup_{(h,b)\in\mathcal{F}}\Big|\sum_{i=1}^n\sigma_ih(\bm{X}_i)\Big|+\frac{16M_0}{n}(1 + C_K\sqrt{\frac{2M_0}{\lambda_n}})\mathbb{E}\Big|\sum_{i=1}^n\sigma_i\Big| \\
&\leq & \frac{16M_0}{\sqrt{n}}\sqrt{\frac{2M_0}{\lambda_n}}C_K+\frac{16M_0}{\sqrt{n}}(1 + C_K\sqrt{\frac{2M_0}{\lambda_n}}) \\
&=& \frac{16M_0}{\sqrt{n}}(1 + 2C_K\sqrt{\frac{2M_0}{\lambda_n}}).
\end{eqnarray*}

Then we have that with probability at least $1-\delta/2$,
\begin{equation}\label{eq:firstterm}
(\mathbb{P}-\mathbb{P}_n)L(h_n,b_n) \leq \frac{16M_0}{\sqrt{n}}(1 + 2C_K\sqrt{\frac{2M_0}{\lambda_n}}) + 2M_0\sqrt{\frac{2\log(4/\delta)}{n}}.
\end{equation}
By the assumption, (\ref{eq:secondterm}), and (\ref{eq:firstterm}), we obtain that with probability at least $1-\delta$,
\begin{equation*}
\mathcal{R}_{T,g}(f_n) - \mathcal{R}^*_{T,g}\leq M_0\sqrt{\frac{2\log(2/\delta)}{n}} + \frac{16M_0}{\sqrt{n}}(1 + 2C_K\sqrt{\frac{2M_0}{\lambda_n}}) + 2M_0\sqrt{\frac{2\log(4/\delta)}{n}} + c\lambda_n^{\beta}.
\end{equation*}
Let $\lambda_n=n^{-\frac{1}{2\beta+1}}$. By Theorem 3.2, we obtain the final result that with probability at least $1-\delta$,
\begin{equation*}
\mathcal{R}(sign(f_n)) - \mathcal{R}^* \leq \tilde{c}\sqrt{\log(4/\delta)}n^{-\frac{\beta}{2\beta+1}}.
\end{equation*}
Here $\tilde{c}=M_0\left(16+3\sqrt{2}+32C_K\sqrt{2M_0}\right)+c$. This completes the proof.
\end{proof}

\textbf{Proof of Lemma 3.9}
\begin{proof}
We first introduce a lemma. It is revised from Lemma 4.1 in \citet{Steinwart:FastRateGaussian} to adapt to our settings for individualized treatment rules.
\begin{lemma} \label{thm:balllemma}
Let $\mathcal{X}$ be the closed unit ball of the Euclidean space $\mathbb{R}^p$, and $P$ be a distribution on $\mathcal{X}\times\mathcal{A}\times\mathcal{M}$ with regular marginal distribution on $\bm{X}$. Recall
$\delta(\bm{x})=\mathbb{E}(R|\bm{X}=\bm{x},A=1) - \mathbb{E}(R|\bm{X}=\bm{x},A=-1)$ for $\bm{x}\in\mathcal{X}$. On $\acute{\mathcal{X}}:=3\mathcal{X}$ we define
\begin{equation*}
\acute{\delta}(\bm{x})=\left\{
\begin{array}{ll}
\delta(\bm{x}) & {\rm if}\: ||\bm{x}||\leq1, \\
\delta\left(\displaystyle \frac{\bm{x}}{||\bm{x}||}\right) & {\rm otherwise},
\end{array}
\right.
\end{equation*}
where $||\cdot||$ is the Euclidean norm.
We also write $\acute{\mathcal{X}}^+=\{\bm{x}\in\acute{\mathcal{X}}: \acute{\delta}(\bm{x})>0\}$,  and $\acute{\mathcal{X}}^-=\{\bm{x}\in\acute{\mathcal{X}}: \acute{\delta}(\bm{x})<0\}$. Finally let $B(\bm{x},r)$ denote the open ball of radius $r$ about $\bm{x}$ in $\mathbb{R}^p$. Then for $\bm{x}\in\mathcal{X}^+$, we have $B(\bm{x},\tau_{\bm{x}})\subset \acute{\mathcal{X}}^+$, and for $\bm{x}\in\mathcal{X}^-$, we have $B(\bm{x},\tau_{\bm{x}})\subset \acute{\mathcal{X}}^-$.
\end{lemma}
The proof is simple, and is the same as that of Lemma 4.1 in \citet{Steinwart:FastRateGaussian}. We omit the proof here. In the lemma, the support is enlarged to ensure that all balls of the form $B(\bm{x},\tau_{\bm{x}})$ are contained in the enlarged support. We return to the proof of Lemma 3.9.

Let ${L}_2(\mathbb{R}^p)$ be the $L_2$-space on $\mathbb{R}^p$ with respect to Lebesque measure, and $\mathcal{H}_{\sigma}(\mathbb{R}^p)$ be the RKHS of the Gaussian RBF kernel $K_{\sigma}$. The linear operator $V_{\sigma}:{L}_2(\mathbb{R}^p)\rightarrow \mathcal{H}_{\sigma}(\mathbb{R}^p)$ defined by
\begin{equation*}
V_{\sigma}\ell(\bm{x}) = \frac{(2\sigma)^{d/2}}{\pi^{d/4}}\int_{\mathbb{R}^p}e^{-2\sigma^2||\bm{x}-\bm{y}||^2}\ell(\bm{y})d\bm{y}, \qquad \ell\in {L}_2(\mathbb{R}^p), \:\bm{x}\in\mathbb{R}^p,
\end{equation*}
is an isometric isomorphism \citep{Steinwart2006:GaussianRBF}. Thus we have,
\begin{equation} \label{eq:al2}
\mathcal{A}(\lambda) \leq \inf_{\ell\in{L}_2(\mathbb{R}^p)}\frac{\lambda}{2}||\ell||^2_{{L}_2(\mathbb{R}^p)}+\mathcal{R}_{T,g}(V_{\sigma}\ell)-\mathcal{R}^*_{T,g}.
\end{equation}
With the notation of Lemma~\ref{thm:balllemma} we fix a measurable $\acute{f}_P:\acute{\mathcal{X}}\rightarrow[-1,1]$ that satisfies $\acute{f}_P=1$ on $\acute{\mathcal{X}}^+$, $\acute{f}_P=-1$ on $\acute{\mathcal{X}}^-$, and $\acute{f}_P=0$ otherwise. For $\ell:=(\sigma^2/\pi)^{p/4}\acute{f}_P$, we immediately obtain,
\begin{equation} \label{eq:L2bound}
||\ell||_{{L}_2(\mathbb{R}^p)} \leq \left(\frac{81\sigma^2}{\pi}\right)^{p/4}\theta(p),
\end{equation}
where $\theta(p)$ denotes the volume of $\mathcal{X}$. As shown in the proof of Theorem 3.2, we have
\begin{equation*}
\mathcal{R}_{T,g}(V_{\sigma}\ell)-\mathcal{R}^*_{T,g} = \mathbb{E}(|\delta(\bm{x})|\cdot|T(V_{\sigma}\ell(\bm{x}))-T(d^*(\bm{x}))|)\leq 2\mathbb{E}(|\delta(\bm{x})|\cdot|V_{\sigma}\ell(\bm{x})-d^*(\bm{x})|).
\end{equation*}
Following the same derivations as in the proof of Theorem 2.7 of \citet{Steinwart:FastRateGaussian}, we also obtain
\begin{equation*}
|V_{\sigma}\ell(\bm{x})-d^*(\bm{x})|\leq 8e^{-\sigma^2\tau_{\bm{x}}^2/(2p)}.
\end{equation*}
The geometric noise assumption yields
\begin{equation} \label{eq:excessriskbound}
\mathcal{R}_{T,g}(V_{\sigma}\ell)-\mathcal{R}^*_{T,g}\leq 16\mathbb{E}(|\delta(\bm{x})|e^{-\sigma^2\tau_{\bm{x}}^2/(2p)}) \leq 16C(2p)^{qp/2}\sigma^{-qp}.
\end{equation}
Combining (\ref{eq:al2}), (\ref{eq:L2bound}) and (\ref{eq:excessriskbound}) yields
\begin{equation*}
\mathcal{A}(\lambda) \leq \left(\frac{81\sigma^2}{\pi}\right)^{p/2}\theta^2(p)\lambda/2+16C(2p)^{qp/2}\sigma^{-qp}.
\end{equation*}
The desired result now follows by taking $\sigma = \lambda^{-\frac{1}{(q+1)p}}$.
\end{proof}

\bibliographystyle{jasa}
\bibliography{myreference}

\begin{thebibliography}{53}
\newcommand{\enquote}[1]{``#1''}
\expandafter\ifx\csname natexlab\endcsname\relax\def\natexlab#1{#1}\fi
\expandafter\ifx\csname url\endcsname\relax
  \def\url#1{{\tt #1}}\fi
\expandafter\ifx\csname urlprefix\endcsname\relax\def\urlprefix{URL }\fi

\bibitem[{Allen(2013)}]{Allen2013:KNIFE}
Allen, G.~I.
\newblock \enquote{Automatic Feature Selection via Weighted Kernels and
  Regularization.}
\newblock {\em Journal of Computational and Graphical Statistics\/},
  22(2):284--299 (2013).

\bibitem[{An and Tao(1997)}]{An1997:DC}
An, L. T.~H. and Tao, P.~D.
\newblock \enquote{Solving a Class of Linearly Constrained Indefinite Quadratic
  Problems by D.C. Algorithms.}
\newblock {\em Journal of Global Optimization\/}, 11(3):253--285 (1997).

\bibitem[{Argyriou et~al.(2006)Argyriou, Hauser, Micchelli, and
  Pontil}]{Argyriou2006:kernelselection}
Argyriou, A., Hauser, R., Micchelli, C.~A., and Pontil, M.
\newblock \enquote{A DC-programming algorithm for kernel selection.}
\newblock In Cohen, W.~W. and Moore, A. (eds.), {\em ICML\/}, volume 148 of
  {\em ACM International Conference Proceeding Series\/}, 41--48. ACM (2006).

\bibitem[{Baron et~al.(2013)Baron, Perrodeau, Boutron, and
  Ravaud}]{Baron2013:multiarm}
Baron, G., Perrodeau, E., Boutron, I., and Ravaud, P.
\newblock \enquote{Reporting of analyses from randomized controlled trials with
  multiple arms: a systematic review.}
\newblock {\em BMC Medicine\/}, 11(1):84 (2013).

\bibitem[{Bartlett and Mendelson(2002)}]{Bartlett:Rademacher2002}
Bartlett, P.~L. and Mendelson, S.
\newblock \enquote{Rademacher and Gaussian Complexities: Risk Bounds and
  Structural Results.}
\newblock {\em Journal of Machine Learning Research\/}, 3:463--482 (2002).

\bibitem[{Breiman(2001)}]{Breiman:RF2001}
Breiman, L.
\newblock \enquote{Random Forests.}
\newblock {\em Machine Learning\/}, 45(1):5--32 (2001).

\bibitem[{Byrd et~al.(1995)Byrd, Lu, Nocedal, and Zhu}]{Byrd1995:LBFGSB}
Byrd, R.~H., Lu, P., Nocedal, J., and Zhu, C.
\newblock \enquote{A Limited Memory Algorithm for Bound Constrained
  Optimization.}
\newblock {\em SIAM Journal on Scientific and Statistical Computing\/},
  16(5):1190--1208 (1995).

\bibitem[{Cohen(1988)}]{Cohen:Power}
Cohen, J.
\newblock {\em Statistical power analysis for the behavioral sciences\/}.
\newblock Hillsdale, NJ: Lawrence Erlbaum Associates, Inc. (1988).

\bibitem[{Collobert et~al.(2006)Collobert, Sinz, Weston, and
  Bottou}]{Collobert2006:RampLoss}
Collobert, R., Sinz, F., Weston, J., and Bottou, L.
\newblock \enquote{Trading convexity for scalability.}
\newblock In {\em Proceedings of the 23rd international conference on Machine
  learning\/}, ICML '06, 201--208. New York, NY, USA: ACM (2006).

\bibitem[{Devroye et~al.(1996)Devroye, Gy{\"{o}}rfi, and
  Lugosi}]{Devroye:PatternRecog1996}
Devroye, L., Gy{\"{o}}rfi, L., and Lugosi, G.
\newblock {\em A Probabilistic Theory of Pattern Recognition\/}.
\newblock Springer (1996).

\bibitem[{Dietterich and Bakiri(1995)}]{Dietterich1995:ECOC}
Dietterich, T.~G. and Bakiri, G.
\newblock \enquote{Solving Multiclass Learning Problems via Error-correcting
  Output Codes.}
\newblock {\em Journal of Artificial Intelligence Research\/}, 2(1):263--286
  (1995).

\bibitem[{Fleming and Harrington(1991)}]{Fleming:Counting1991}
Fleming, T. and Harrington, D.
\newblock {\em Counting Processes and Survival Analysis\/}.
\newblock Wiley (1991).

\bibitem[{Fung and Mangasarian(2004)}]{Fung2004:L1SVM}
Fung, G.~M. and Mangasarian, O.~L.
\newblock \enquote{A Feature Selection Newton Method for Support Vector Machine
  Classification.}
\newblock {\em Computational Optimization and Applications.\/}, 28(2):185--202
  (2004).

\bibitem[{Grandvalet and Canu(2002)}]{Grandvalet2002:FeatureSelection}
Grandvalet, Y. and Canu, S.
\newblock \enquote{Adaptive Scaling for Feature Selection in SVMs.}
\newblock In Becker, S., Thrun, S., and Obermayer, K. (eds.), {\em Advances in
  Neural Information Processing System\/}, 553--560. MIT Press (2002).

\bibitem[{Gunter et~al.(2011)Gunter, Zhu, and Murphy}]{Gunter2011:Qualitative}
Gunter, L., Zhu, J., and Murphy, S.
\newblock \enquote{Variable Selection for Qualitative Interactions.}
\newblock {\em Statistical Methodology\/}, 8:42--55 (2011).

\bibitem[{Hastie et~al.(2001)Hastie, Tibshirani, and Friedman}]{Hastie2001}
Hastie, T., Tibshirani, R., and Friedman, J.~H.
\newblock {\em The Elements of Statistical Learning\/}.
\newblock Springer (2001).

\bibitem[{Havlir et~al.(2011)Havlir, Kendall, Ive, Kumwenda, Swindells, Qasba,
  Luetkemeyer, Hogg, Rooney, Wu, Hosseinipour, Lalloo, Veloso, Some,
  Kumarasamy, Padayatchi, Santos, Reid, Hakim, Mohapi, Mugyenyi, Sanchez, Lama,
  Pape, Sanchez, Asmelash, Moko, Sawe, Andersen, and
  Sanne}]{Havlir:ARTTiming2011}
Havlir, D.~V., Kendall, M.~A., Ive, P., Kumwenda, J., Swindells, S., Qasba,
  S.~S., Luetkemeyer, A.~F., Hogg, E., Rooney, J.~F., Wu, X., Hosseinipour,
  M.~C., Lalloo, U., Veloso, V.~G., Some, F.~F., Kumarasamy, N., Padayatchi,
  N., Santos, B.~R., Reid, S., Hakim, J., Mohapi, L., Mugyenyi, P., Sanchez,
  J., Lama, J.~R., Pape, J.~W., Sanchez, A., Asmelash, A., Moko, E., Sawe, F.,
  Andersen, J., and Sanne, I.
\newblock \enquote{Timing of Antiretroviral Therapy for HIV-1 Infection and
  Tuberculosis.}
\newblock {\em New England Journal of Medicine\/}, 365(16):1482--1491 (2011).

\bibitem[{Hsu et~al.(2003)Hsu, Chang, and Lin}]{Hsu2003:svmguide}
Hsu, C.-W., Chang, C.-C., and Lin, C.-J.
\newblock \enquote{A practical guide to support vector classification.}
\newblock Technical report, Department of Computer Science, National Taiwan
  University, Taipei 106, Taiwan (2003).

\bibitem[{Kimeldorf and Wahba(1971)}]{Kimeldorf:Representer}
Kimeldorf, G. and Wahba, G.
\newblock \enquote{Some results on Tchebycheffian spline functions.}
\newblock {\em Journal of Mathematical analysis and applications\/}, 33:82--95
  (1971).

\bibitem[{Lafferty and Wasserman(2008)}]{Lafferty2008:Rodeo}
Lafferty, J. and Wasserman, L.
\newblock \enquote{Rodeo: Sparse, greedy nonparametric regression.}
\newblock {\em The Annals of Statistics\/}, 36(1):28--63 (2008).

\bibitem[{Ledoux and Talagrand(1991)}]{Ledoux:Banach1991}
Ledoux, M. and Talagrand, M.
\newblock {\em Probability in Banach Spaces: isoperimetry and processes\/}.
\newblock Springer (1991).

\bibitem[{Lee et~al.(2004)Lee, Lin, and Wahba}]{Lee2004:MSVM}
Lee, Y., Lin, Y., and Wahba, G.
\newblock \enquote{Multicategory Support Vector Machines, theory, and
  application to the classification of microarray data and satellite radiance
  data.}
\newblock {\em Journal of the American Statistical Association\/}, 99:67--81
  (2004).

\bibitem[{Lin and Zhang(2006)}]{Lin2006:COSSO}
Lin, Y. and Zhang, H.~H.
\newblock \enquote{Component selection and smoothing in multivariate
  nonparametric regression.}
\newblock {\em The Annals of Statistics\/}, 34(5):2272--2297 (2006).

\bibitem[{Morales and Nocedal(2011)}]{Morales2011:LBFGSB}
Morales, J.~L. and Nocedal, J.
\newblock \enquote{Remark on `Algorithm 778: L-BFGS-B: Fortran Subroutines for
  Large-scale Bound Constrained Optimization'.}
\newblock {\em ACM Transactions on Mathematical Software\/}, 38(1):7:1--7:4
  (2011).

\bibitem[{Murphy(2003)}]{Murphy:OptimalDTR2003}
Murphy, S.~A.
\newblock \enquote{Optimal dynamic treatment regimes.}
\newblock {\em Journal of the Royal Statistical Society: Series B\/},
  65(2):331--355 (2003).

\bibitem[{Murphy(2005)}]{Murphy:Design2005}
---.
\newblock \enquote{An experimental design for the development of adaptive
  treatment strategies.}
\newblock {\em Statistics in medicine\/}, 24(10) (2005).

\bibitem[{Nocedal(1980)}]{Nocedal1980:LBFGS}
Nocedal, J.
\newblock \enquote{Updating {Quasi-Newton} Matrices with Limited Storage.}
\newblock {\em Mathematics of Computation\/}, 35(151):773--782 (1980).

\bibitem[{Nocedal and Wright(2006)}]{Nocedal2006:NO}
Nocedal, J. and Wright, S.~J.
\newblock {\em Numerical Optimization\/}.
\newblock New York: Springer, 2nd edition (2006).

\bibitem[{Qian and Murphy(2011)}]{Qian:ITR2011}
Qian, M. and Murphy, S.~A.
\newblock \enquote{Performance guarantees for individualized treatment rules.}
\newblock {\em The Annals of Statistics\/}, 39(2):1180--1210 (2011).

\bibitem[{Rakotomamonjy(2003)}]{Rakotomamonjy2003:svm}
Rakotomamonjy, A.
\newblock \enquote{Variable selection using svm based criteria.}
\newblock {\em J. Mach. Learn. Res.\/}, 3:1357--1370 (2003).

\bibitem[{Robins(2004)}]{Robins2004:SNM}
Robins, J.~M.
\newblock \enquote{Optimal Structural Nested Models for Optimal Sequential
  Decisions.}
\newblock In Lin, D. and Heagerty, P. (eds.), {\em Proceedings of the Second
  Seattle Symposium in Biostatistics\/}, volume 179 of {\em Lecture Notes in
  Statistics\/}, 189--326. Springer New York (2004).

\bibitem[{Rudin(1987)}]{Rudin:Measure1987}
Rudin, W.
\newblock {\em Real and Complex Analysis, 3rd Ed.\/}.
\newblock New York, NY, USA: McGraw-Hill, Inc. (1987).

\bibitem[{Schmidt(2010)}]{Schmidt2010:LASSO}
Schmidt, M.
\newblock \enquote{Graphical model structure learning with l1-regularization.}
\newblock Ph.D. thesis, The University of British Columbia (2010).

\bibitem[{Steinwart and Christmann(2008)}]{Steinwart:SVM2008}
Steinwart, I. and Christmann, A.
\newblock {\em Support Vector Machines\/}.
\newblock Springer (2008).

\bibitem[{Steinwart et~al.(2006)Steinwart, Hush, and
  Scovel}]{Steinwart2006:GaussianRBF}
Steinwart, I., Hush, D.~R., and Scovel, C.
\newblock \enquote{An Explicit Description of the Reproducing Kernel Hilbert
  Spaces of Gaussian RBF Kernels.}
\newblock {\em IEEE Transactions on Information Theory\/}, 52(10):4635--4643
  (2006).

\bibitem[{Steinwart and Scovel(2007)}]{Steinwart:FastRateGaussian}
Steinwart, I. and Scovel, C.
\newblock \enquote{Fast rates for support vector machines using Gaussian
  kernels.}
\newblock {\em The Annals of Statistics\/}, 35(2):575--607 (2007).

\bibitem[{Therneau et~al.(1990)Therneau, Grambsch, and
  Fleming}]{Therneau:MResid1990}
Therneau, T.~M., Grambsch, P.~M., and Fleming, T.~R.
\newblock \enquote{Martingale-based residuals for survival models.}
\newblock {\em Biometrika\/}, 77(1):147--160 (1990).

\bibitem[{Tibshirani(1994)}]{Tibshirani1994:lasso}
Tibshirani, R.
\newblock \enquote{Regression Shrinkage and Selection Via the Lasso.}
\newblock {\em Journal of the Royal Statistical Society, Series B\/},
  58:267--288 (1994).

\bibitem[{Treggiari et~al.(2011)Treggiari, Retsch-Bogart, Mayer-Hamblett, Khan,
  Kulich, Kronmal, Williams, Hiatt, Gibson, Spencer, Orenstein, Chatfield,
  Froh, Burns, Rosenfeld, and Ramsey}]{Treggiari2011:EPIC}
Treggiari, M.~M., Retsch-Bogart, G., Mayer-Hamblett, N., Khan, U., Kulich, M.,
  Kronmal, R., Williams, J., Hiatt, P., Gibson, R.~L., Spencer, T., Orenstein,
  D., Chatfield, B.~A., Froh, D.~K., Burns, J.~L., Rosenfeld, M., and Ramsey,
  B.~W.
\newblock \enquote{Comparative efficacy and safety of 4 randomized regimens to
  treat early pseudomonas aeruginosa infection in children with cystic
  fibrosis.}
\newblock {\em Archives of Pediatrics \& Adolescent Medicine\/},
  165(9):847--856 (2011).

\bibitem[{Treggiari et~al.(2009)Treggiari, Rosenfeld, Mayer-Hamblett,
  Retsch-Bogart, Gibson, Williams, Emerson, Kronmal, and
  Ramsey}]{Treggiari2009:EPIC}
Treggiari, M.~M., Rosenfeld, M., Mayer-Hamblett, N., Retsch-Bogart, G., Gibson,
  R.~L., Williams, J., Emerson, J., Kronmal, R.~A., and Ramsey, B.~W.
\newblock \enquote{Early anti-pseudomonal acquisition in young patients with
  cystic fibrosis: Rationale and design of the {EPIC} clinical trial and
  observational study.}
\newblock {\em Contemporary Clinical Trials\/}, 30(3):256 -- 268 (2009).

\bibitem[{Vapnik(1998)}]{Vapnik:svm98}
Vapnik, V.~N.
\newblock {\em Statistical Learning Theory\/}.
\newblock New York: John Wiley and Sons (1998).

\bibitem[{Wang et~al.(2008)Wang, Zhu, and Zou}]{Wang2008:hhsvm}
Wang, L., Zhu, J., and Zou, H.
\newblock \enquote{Hybrid huberized support vector machines for microarray
  classification and gene selection.}
\newblock {\em Bioinformatics\/}, 24(3):412--419 (2008).

\bibitem[{Weston et~al.(2000)Weston, Mukherjee, Chapelle, Pontil, Poggio, and
  Vapnik}]{Weston2000:SVMSelection}
Weston, J., Mukherjee, S., Chapelle, O., Pontil, M., Poggio, T., and Vapnik, V.
\newblock \enquote{Feature Selection for {SVMs}.}
\newblock In {\em Advances in Neural Information Processing System\/},
  668--674. MIT Press (2000).

\bibitem[{Wu and Liu(2007)}]{Wu2007:RobustSVM}
Wu, Y. and Liu, Y.
\newblock \enquote{Robust truncated-hinge-loss support vector machines.}
\newblock {\em Journal of the American Statistical Association\/}, 102:974--983
  (2007).

\bibitem[{Yuille and Rangarajan(2003)}]{Yuille2003:CCCP}
Yuille, A.~L. and Rangarajan, A.
\newblock \enquote{The concave-convex procedure.}
\newblock {\em Neural Computation\/}, 15(4):915--936 (2003).

\bibitem[{Zhang et~al.(2012)Zhang, Tsiatis, Laber, and
  Davidian}]{Zhang:RobustITR2012}
Zhang, B., Tsiatis, A.~A., Laber, E.~B., and Davidian, M.
\newblock \enquote{A Robust Method for Estimating Optimal Treatment Regimes.}
\newblock {\em Biometrics\/}, 68(4):1010--1018 (2012).

\bibitem[{Zhao et~al.(2009)Zhao, Kosorok, and Zeng}]{Zhao:RL2009}
Zhao, Y., Kosorok, M.~R., and Zeng, D.
\newblock \enquote{Reinforcement learning design for cancer clinical trials.}
\newblock {\em Statistics in Medicine\/}, 28(26):3294--3315 (2009).

\bibitem[{Zhao et~al.(2012)Zhao, Zeng, Rush, and Kosorok}]{Zhao:OWL2012}
Zhao, Y., Zeng, D., Rush, A.~J., and Kosorok, M.~R.
\newblock \enquote{Estimating Individualized Treatment Rules Using Outcome
  Weighted Learning.}
\newblock {\em Journal of the American Statistical Association\/},
  107(499):1106--1118 (2012).

\bibitem[{Zhao et~al.(2011)Zhao, Zeng, Socinski, and
  Kosorok}]{Zhao2011:Reinforcement}
Zhao, Y., Zeng, D., Socinski, M.~A., and Kosorok, M.~R.
\newblock \enquote{Reinforcement Learning Strategies for Clinical Trials in
  Nonsmall Cell Lung Cancer.}
\newblock {\em Biometrics\/}, 67(4):1422--1433 (2011).

\bibitem[{Zhao et~al.(2014)Zhao, Zeng, Laber, and Kosorok}]{Zhao2014:DTR}
Zhao, Y.-Q., Zeng, D., Laber, E.~B., and Kosorok, M.~R.
\newblock \enquote{New Statistical Learning Methods for Estimating Optimal
  Dynamic Treatment Regimes.}
\newblock {\em Journal of the American Statistical Association\/}, In press
  (2014).

\bibitem[{Zhu and Hastie(2004)}]{Zhu2004:PLR}
Zhu, J. and Hastie, T.
\newblock \enquote{Classification of gene microarrays by penalized logistic
  regression.}
\newblock {\em Biostatistics\/}, 5(3):427--443 (2004).

\bibitem[{Zhu et~al.(2003)Zhu, Rosset, Hastie, and Tibshirani}]{Zhu2003:L1SVM}
Zhu, J., Rosset, S., Hastie, T., and Tibshirani, R.
\newblock \enquote{1-Norm Support Vector Machines.}
\newblock In {\em Advances in Neural Information Processing System\/} (2003).

\bibitem[{Zou and Hastie(2005)}]{Zou2005:elasticnet}
Zou, H. and Hastie, T.
\newblock \enquote{Regularization and variable selection via the Elastic Net.}
\newblock {\em Journal of the Royal Statistical Society, Series B\/},
  67:301--320 (2005).

\end{thebibliography}

\end{document}